\documentstyle[11pt,amssymb,epsfig,amsmath]{article}

\begin{document}
\newcommand{\be}{\begin{equation}}
\newcommand{\ee}{\end{equation}}
\newcommand{\ba}{\begin{eqnarray}}
\newcommand{\ea}{\end{eqnarray}}
\newcommand{\no}{\nonumber \\}
\newcommand{\gsim}{\mathrel{\hbox{\rlap{\lower.55ex \hbox {$\sim$}}
                   \kern-.3em \raise.4ex \hbox{$>$}}}}
\newcommand{\lsim}{\mathrel{\hbox{\rlap{\lower.55ex \hbox {$\sim$}}
                   \kern-.3em \raise.4ex \hbox{$<$}}}}

\def\be{\begin{eqnarray}}
\def\ee{\end{eqnarray}}
\def\bea{\be}
\def\eea{\ee}
\newcommand{\e}{{\mbox{e}}}
\def\del{\partial}
\def\vr{{\vec r}}
\def\vk{{\vec k}}
\def\vq{{\vec q}}
\def\vp{{\vec p}}
\def\vP{{\vec P}}
\def\vt{{\vec \tau}}
\def\vs{{\vec \sigma}}
\def\vJ{{\vec J}}
\def\vB{{\vec B}}
\def\hatr{{\hat r}}
\def\hatk{{\hat k}}
\def\roughly#1{\mathrel{\raise.3ex\hbox{$#1$\kern-.75em%
\lower1ex\hbox{$\sim$}}}}
\def\lsim{\roughly<}
\def\gsim{\roughly>}
\def\fm{{\mbox{fm}}}
\def\vx{{\vec x}}
\def\vy{{\vec y}}
\def\({\left(}
\def\){\right)}
\def\[{\left[}
\def\]{\right]}
\def\EM{{\rm EM}}
\def\barp{{\bar p}}
\def\zz{{z \bar z}}
\def\mus{{\cal M}_s}
\def\abs#1{{\left| #1 \right|}}
\def\ve{{\vec \epsilon}}
\def\nlo#1{{\mbox{N$^{#1}$LO}}}
\def\MS{{\mbox{M1V}}}
\def\mut{{\mbox{M1S}}}
\def\Qt{{\mbox{E2S}}}
\def\rM{{\cal R}_{\rm M1}}\def\rE{{\cal R}_{\rm E2}}
\def\la{{\Big<}}
\def\ra{{\Big>}}
\def\lsim{\mathrel{\rlap{\lower3pt\hbox{\hskip1pt$\sim$}}
     \raise1pt\hbox{$<$}}} 
\def\gsim{\mathrel{\rlap{\lower3pt\hbox{\hskip1pt$\sim$}}
     \raise1pt\hbox{$>$}}} 
\def\N{${\cal N}\,\,$}

\def\ka{{\kappa}}
\def\lam{{\lambda}}
\def\dlt{{\delta}}
\def\omg{{\omega}}
\def\lv{\lvert}
\def\rv{\rvert}
\def\wh{\tilde h}

\def\J#1#2#3#4{ {#1} {\bf #2} (#4) {#3}. }
\def\PRL{Phys. Rev. Lett.}
\def\PL{Phys. Lett.}
\def\PLB{Phys. Lett. B}
\def\NP{Nucl. Phys.}
\def\NPA{Nucl. Phys. A}
\def\NPB{Nucl. Phys. B}
\def\PR{Phys. Rev.}
\def\PRC{Phys. Rev. C}

\renewcommand{\thefootnote}{\arabic{footnote}}
\setcounter{footnote}{0}

\vskip 0.4cm \hfill { }
 \hfill {\today} \vskip 1cm

\begin{center}
{\LARGE\bf  Toward the AdS/CFT Gravity Dual for High Energy
  Collisions. 3.Gravitationally Collapsing Shell and Quasiequilibrium
   }
\date{\today}

\vskip 1cm {\large Shu
Lin\footnote{E-mail:slin@grad.physics.sunysb.edu},
and Edward
Shuryak\footnote{E-mail:shuryak@tonic.physics.sunysb.edu}
 }


\end{center}

\vskip 0.5cm

\begin{center}

 {\it Department of Physics and Astronomy, SUNY Stony-Brook,
NY 11794}

\end{center}

\vskip 0.5cm

\begin{abstract}
  The equilibration of matter and onset of hydrodynamics
can be understood in the AdS/CFT context as 
a gravitational collapse process, in which ``collision debris''
create a horizon. In this paper we consider
the simplest geometry possible, a flat shell (or membrane) falling
in the holographic direction toward the horizon. The metric is
a combination of two well known solutions: thermal AdS above the 
shell and pure AdS below, while motion of the shell is given by
the Israel junction condition. Furthermore,
when the shell motion can be considered slow,
we were able to solve for two-point functions of all boundary stress tensor
and found that an observer on the boundary sees a very peculiar
$quasiequilibrium$: while the average 
stress tensor $<T_{\mu\nu>}$ contains the equilibrium
plasma energy and pressure at all times, the
 spectral densities of the correlators
(related with occupation probabilities of the modes)
reveal additional oscillating
terms absent in equilibrium. This is explained by the ``echo''
    phenomenon, a partial return of the field coherence 
at certain  ``echo'' times.
\end{abstract}

\newpage

\renewcommand{\thefootnote}{\#\arabic{footnote}}
\setcounter{footnote}{0}

\section{Introduction }

Observation at RHIC of collective flows in relativistic heavy ion collisions,
well described by ideal hydrodynamics \cite{shuryak,heinz}
have lead to a paradigm shift in the field toward studies of strongly
coupled plasmas. AdS/CFT correspondence \cite{adscft} is one of actively
pursuit directions, both for understanding of strongly coupled
gauge theories in general and properties of Quark-Gluon Plasma (QGP)
in particularly: for recent review see \cite{Shuryak:2008eq}.

This paper continues the line of work of our two previous papers,
\cite{Lin:2006rf} and 
  \cite{Lin:2007fa}, which we will call I and II respectively below.
Those papers include extensive introduction explaining
 our approach to the problem, which we will
not repeat here. Let us just say that in those works we dealt with
``elementary'' collisions -- calculating the shape of the falling
string between two departing charges and its hologram at the boundary 
 -- which in the QCD language can be related to $e^+e^-$ annihilation
into a pair of heavy quarks or $pp$ collisions. Thus 
in I and II there was no 
horizon on the gravity side and  no temperature or entropy on the
matter side.
In this paper we address these issues, related with heavy ion 
collisions and equilibration.

Properties of equilibrium strongly  coupled conformal
 plasma is by now well studied
in significant detail in the AdS setting, in static thermal AdS-BH
 metric
suggested by
Witten. Many details about quasinormal modes of this metric and in
 particularly the correlators of stress tensors and their 
spectral densities are known, we especially recommend
\cite{qnm,kovtun,starinets,nunez}. Recent developments included 
flowing near-equilibrium state, with slowly/gradually
deformed horizons and derivation of hydrodynamics up to
second order in gradients: we will not use those here,
see references in a review \cite{Shuryak:2008eq}.

The most challenging task of the theory now is
the understanding of how matter manages to equilibrate
 so rapidly in RHIC 
collisions, and what exactly such equilibration
means microscopically. The success of hydrodynamics
in describing RHIC elliptic flow data
  seems to suggest the thermalization time of the order of 
$0.5 \, fm/c$, 
yet its mechanism remains unclear.
The quest for its  mechanism involves studies of
various phenomenology and theoretical approaches.
We will not attempt to review it and just mention one 
approach to ideas of which we will
refer below, the ``Glasma model'': see Ref.\cite{raju} and references therein. 
It is based on classical Yang-Mills equations and starts with
 the so called Color Glass Condensate
 initial condition.
 While the coupling is assumed weak in this approach,
strong coherent fields make its behavior  nonperturbative.
 
Our approach also attempts to address the transition from
 Glasma-like  initial coherent gauge fields 
 to incoherent near-thermal QGP, but in strong coupling and
thus 
 the AdS/CFT setting. Since the vacuum corresponds to extremal
black hole solution -- pure AdS geometry without a horizon -- while
the thermal field theory is dual to
 AdS black hole with a horizon, the main issue is 
how deposition of an extra mass --
from ``debris'' of the collision -- leads to
excitation of the extremal to non-extremal black hole and
dynamical
creation of the horizon.
In simpler words, we have to follow 
 some kind of a gravitational collapse.

In I and II we already discuss qualitatively how falling debris
created in the collisions -- e.g. large number of falling strings
between departing charges -- make a falling matter shell or membrane.
While in I we studied its fall ignoring its own weight,
we have to include it now, as its near-horizon
 breaking is entirely due to
adding the mass of the shell to total
gravity. We have argued in those papers that in principle one can
address the problem by following the motion of {\em two membranes},
the  $shell$ and that describing $horizon$ (a la ``membrane paradigm''
see book by Thorne and collaborators \cite{Thorne:1986iy}). 

For now we do not study motion of those two membranes in
realistic geometry, for obvious reasons starting with
the simplest geometry possible. We assume that the matter shell
(and thus the horizon) is $flat$ -- that is independent on our world
3 spatial coordinates, and moves only in the 5-th
holographic direction\footnote{the equilibration in this setting is not
due to spatial gradient as in hydrodynamics}. The setup of this
gravitational collapse model is described in Sec.\ref{shell}.
The equation of motion  for the shell is given by 
Israel junction condition\cite{israel}
which we solve numerically. We find how its trajectory depends on the
property of the shell; but in all cases the distant observer
sees its slow approach to the horizon at late time. 

 Early works along this path include important papers
\cite{shellon,danielsson,giddings} which we partly follow.
They consider a collapsing shell geometry, but unlike
us they use as a probe some bulk scalar field and we don't find
their boundary conditions at the shell sufficiently convincing,
and we tried to improve those by using gravitons instead. 

 What the observer sees as the shell falls is dual to
the evolution of ${\cal N}=4$ SYM from certain initial ensemble to
the final
thermal equilibrium. The insight into the problem
of thermalization is thus obtained
by studying various observables --
 the induced stress tensor and its correlation functions
 on the boundary -- while the shell is somewhere in its process
of falling. The former is given by one-point function
and the metric above the shell, which in our geometry is
time $independent$ AdS black hole metric. Thus the ``single point 
observer''
who is only able to measure the {\em average density and pressure}
 would be
driven to the conclusion that the matter is fully 
equilibrated at all times. More sophisticated  ``two point 
observer''-- measuring the stress tensor correlation functions --
would however be able to see the deviations 
the from the thermal ensemble. We compute those
deviations in Sec.
 \ref{correlation},  using various components of the graviton 
perturbations to probe the gravitational background with a shell.

As a significant technical advance of this work, we show that
 unique prescription for boundary conditions for the gravity
waves on the shell 
follows from the junction
condition itself. Thus we show how correctly
propagate the graviton
wave  across the shell, relating the obvious infalling
conditions near the AdS center to what is seen on the boundary.
Explicit expressions are obtained for two-point function
in near equilibrium stages, when the shell is
close to the horizon. We solve the wave equations both numerically
and using the WKB approximation, finding good agreement between
 the two. 
 Possible implications of the results are finally summarized
and discussed in 
Sec.\ref{conclude}.

\section{Gravitationally collapsing shell in AdS}\label{shell}
\subsection{The background metric }
Our setting includes the basic AdS background, described by
the metric  $ds^2=\frac{-dt^2+d{\vec x}^2+dz^2}{z^2}$. Its
 holographic coordinate $z$ is zero at the boundary (UV) and
infinity at the AdS center (IR).
 The problem considered is a simple generalization of
Israel's original problem, which was 
collapsing spherical shell in asymptotically flat 3d space.
And it shares its main feature: although the shell
is falling with its radial position  time depending,
the gravity both inside and outside it is time independent. 
Furthermore,  inside a sphere there is no
 influence of the shell's existence at all:
the famous statement  going back to Newton
himself. The gravity outside only knows the total shell mass.

It is not difficult to prove that the same is
true for flat shell in the AdS setting as well.
Starting with a generic form:
\be\label{metric}
ds^2=-A(z,t)dt^2+B(z,t)d{\vec x}^2+C(z,t)dz^2
\ee
one can set $B=\frac{1}{z^2}$ by a coordinate transformation.
The metric has to satisfy the vacuum Einstein equation
\be\label{einstein}
G_{\mu\nu}-\Lambda g_{\mu\nu}=0
\ee
with $\Lambda=6$  both above\footnote{The reader is reminded
that the coordinate $z$ is
inversely proportional to radial coordinate, thus somewhat
counterintuitive inequalities. }
$z<z_m$ and 
below $z>z_m$ the shell's position $z_m$. We will also refer to
those as ``outside'' and ``inside'' regions below.

The $tz$ component tells us that $\del_t C=0$. The $tt$,$zz$ equations
 are used to
obtain:
\be
&&C(z,t)=\frac{1}{z^2(1+kz^4)} \no
&&A(z,t)=F(t)\frac{1+kz^4}{z^2}
\ee
$F(t)$ can be dropped by a rescaling in the $t$ coordinate.
Now we require the metric should reduce to the AdS form 
 infinitely far away from the shell.
as $z\rightarrow0$, we can have $k=-\frac{1}{z_h^4}$, then outside the shell
the metric is in form of translationally invariant
AdS-black hole(AdS-BH). On the other hand, inside
at the AdS center $z\rightarrow\infty$
we have to set $k=0$ to suppress the $z^4$ term.
 Therefore the inside is just
AdS metric.

So the background metric is the combination of AdS-BH and AdS, with the
two metrics separated by a shell. We will use the metric
in the usual form
\be\label{eq:metric}
ds^2=\frac{R^2}{z^2}(-f(z)dt^2+d{\vec x}^2+dz^2/f(z))
\ee
with $f=1-\frac{z^4}{z_h^4}$ (or $f=1$)  outside (or inside)
 the shell position $z_m$.
${\vec x}$ and $z$ are both continuous in order for $\frac{d{\vec x}^2}{z^2}$
 to match.

Note that a singularity at $z=z_h$ is outside the region where
the former metric is used, as $z_m<z_h$. This does not mean
that there is no horizon in the problem: in spite of pure AdS
metric inside, the dynamical horizon and trapped surface (both time
dependent) do exists.

\subsection{ Israel junction conditions and the falling shell}\label{jc}
As in \cite{Lin:2006rf}, the strings in AdS bulk were modeled by a shell
(membrane), the action of which is given by:
\be
S_m=-p\int d^4\sigma\sqrt{-detg_{ij}}
\ee
where $g_{ij}$ is the induced metric on the shell. $p$ is the only parameter
characterizing the shell.

We use Lanczos equation to study the falling of the shell:

\be\label{eq:lanczos}
[K_{ij}]-g_{ij}[K]=-\ka_5^2 S_{ij}
\ee
where $[K_{ij}]=K_{ij}^+-K_{ij}^-$ 
and $K_{ij}=n_\alpha(\frac{\del^2x^\alpha}{\del\xi^i\del\xi^j}
+\Gamma^\alpha_{\beta\gamma}\frac{\del x^\beta}{\del\xi^i}\frac{\del x^\gamma}{\del\xi^j})$

We parametrize the induced metric on the shell as follows:
\be
ds^2_\Sigma=-\frac{d\tau^2}{z^2}+\frac{d\vec{x}^2}{z^2}
\ee
We choose $\alpha=t,z,\vec{x}$ and $i=\tau,\vec{x}$
Assume the EOM is given by $z(\tau),t(\tau)$.

The norm $n_{\alpha}$ is determined from the condition $n_{\alpha}dx^\alpha=0$ and
$n^2=1$ as:
\be
n_{\alpha}=(-\frac{\dot z}{z},\vec{0},\frac{\dot{t}}{z})
\ee
Note here the norm points to the AdS center($z=\infty$). Therefore we have 
$+$:inside; $-$:outside

The curvature K is calculated as follows:
\be
&&K_{\tau\tau}=\frac{\dot{t}}{z}(\frac{ff'+2f\ddot{z}}{2(f+\dot{z}^2)}-\frac{f}{z})\\
&&K_{xx}=\frac{\dot{t}}{z}\frac{f}{z}
\ee

$S_{ij}$ is determined from the shell action:
$S_{ij}=\frac{2}{\sqrt{-g}}\frac{\delta S_m}{\delta g^{ij}}=pg_{ij}$. 
(\ref{eq:lanczos}) becomes:
\be
&&[K_{\tau\tau}-g_{\tau\tau}K]=\frac{\ka_5^2 p}{z^2}\label{eq:1} \\
&&[K_{xx}-g_{xx}K]=-\frac{\ka_5^2 p}{z^2}\label{eq:2}
\ee

\be
&&(\ref{eq:1}) \Rightarrow \sqrt{1+\dot{z}^2}-\sqrt{f+\dot{z}^2}
=\frac{\ka_5^2 p}{3} \label{eq:3}\\
&&(\ref{eq:1})+(\ref{eq:2}) \Rightarrow [\frac{\dot{t}}{z}\frac{ff'+2f\ddot{z}}{2(f+\dot{z}^2)}]=0 \Rightarrow [\frac{(f'+2\ddot{z})\dot{z}}{2(f+\dot{z}^2)}]=0
\Rightarrow [\frac{d\sqrt{f+\dot{z}^2}}{d\tau}]=0 \label{eq:4}
\ee

(\ref{eq:4}) is solved by (\ref{eq:3}) with integration constant determined!

\be
\dot{z}=\sqrt{(\frac{\ka_5^2 p}{6})^2+(\frac{3}{2\ka_5^2 p})^2(1-f)^2-\frac{1+f}{2}}
\ee

The falling velocity seen by distant observer is given by:

\be\label{trajectory}
\frac{dz}{dt}=\frac{\dot{z}}{\dot{t}}
=\frac{f\sqrt{(\frac{\ka_5^2 p}{6})^2+(\frac{3}{2\ka_5^2 p})^2(1-f)^2-\frac{1+f}{2}}}{\frac{\ka_5^2 p}{6}+\frac{3}{2\ka_5^2 p}(1-f)}
\ee

Suppose the shell starts falling at $z=z_0>0$ with vanishing initial velocity.
The horizon radius $z_h$ can be expressed in terms of $z_0$ and $\ka_5^2 p$:

\be\label{eq:5}
\frac{z_0^4}{z_h^4}=1-f(z_0)=4(1-\frac{\ka_5^2 p}{6})\frac{\ka_5^2 p}{6}
\ee

Note (\ref{eq:5}) implies another constraint $\ka_5^2 p<6$. The independent 
parameters $z_0$ and $p$ should be estimated from the initial conditions(e.g.
energy density, particle number), these will determine the equilibrium
temperature of the evolution.

\begin{figure}[t]
\includegraphics[width=0.5\textwidth]{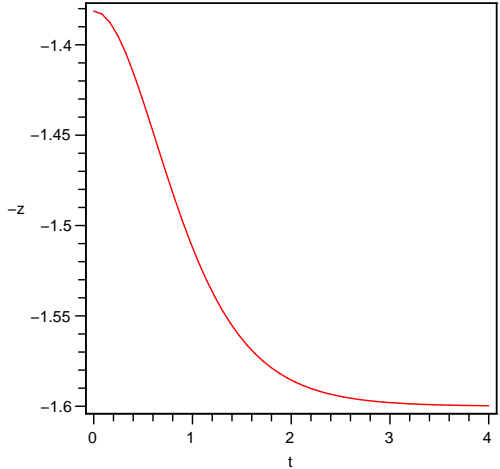}
\caption{\label{stages}The shell trajectory as a function of time.
It starts at rest at $z=z_0$
with a constant acceleration, followed by a constant falling and 
eventually approaches the horizon in a exponential fashion. 
The parameter we choose are $\ka_5^2 p=1$ and $z_h=1.6$}
\end{figure}

With chosen parameters, it is easy to integrate (\ref{trajectory}) to
 give the trajectory of the shell.
We have plotted the shell trajectory in Fig.\ref{stages}. It shows 
three stages of falling, initial acceleration ($z=z_0+\#t^2$),
 intermediate near-constant velocity fall, and the final near-horizon
freezing with exponentially small deviation  ($z=z_h-e^{-\#t}$).

Finally, what are the physical meaning of the parameters
of our model, the
initial position $z_0$ and the shell tension $p$?
The wave functions of the colliding nuclei are believed to
be \cite{CGC} concentrated at the certain momentum scale called the
``saturation scale'' $Q_s$, which depends on collision energy
and nuclei: it is about 1.5 GeV at RHIC.
 The holographic coordinate has the meaning of
the inverse momentum scale in the renormalization group sense,
thus its initial value should be identified with inverses saturation scale
$z_0=1/Q_s$. Further ``falling'' corresponds to motion of the
wave function into the infrared direction, till equilibrium is
reached. The initial temperatures at RHIC 
(and thus $z_h=1/\pi T$ are believed to be about .35 $GeV$,
with $\pi T\approx 1.1\, GeV$: thus the expected inequality $z_h>z_0$
is satisfied. 

(Physical fireballs not only equilibrate but also expand,
with $T$ decreasing by a factor 3-4 in RHIC collisions. This
 would correspond to a $departing$ horizon toward 
larger $z$, as suggested in \cite{Shuryak:2005ia}, but
this feature is  of course not included in the present model.)

The corresponding tension of the shell $p$ may be calculated 
from (\ref{eq:5}). Although we do not follow this direction here, 
one may attempt to calculate it in a particular
model of the collisions. In particular, the so called Lund
model prescribe how many color strings per transverse area
is created, and as the shell is but an approximation
to all those strings falling together, its tension may
be identified with the $sum$ of the tensions of all the strings.

\section{Correlation functions of the Gauge Theory}\label{correlation}

With the complete gravity background at hand, we are ready to study the
property of the dual gauge theory under thermalization. 
We already commented that since the metric is asymptoticly AdS-BH, the 
one-point function of the stress tensor is the same as thermal SYM
result. Therefore we consider the two-point functions as the
expected place to find deviations from the equilibrated thermal ensemble.

The most standard
way is to study a probe field in the background which is dual to some 
boundary operator, then use AdS/CFT prescription to find the correlation 
function. The simplest probe field used in early works was the bulk
 scalar
\cite{shellon,danielsson,giddings}.  However we choose
to use various components of 
the metric field, $h_{mn}(m,n=t,{\vec x})$, which is dual to the boundary 
stress tensor. One obvious reason is to avoid the introduction
of new field into the problem. The other reason involves the matching condition
of $h_{mn}$ on the shell, as will become clear later. 
Also we only probe 
the geometry after the creation of the shell.

Thus we  solve for the metric perturbation 
$h_{mn}(t,x)$ propagating in the bulk specified above.
 This is a very difficult task in general, because the 
shell is always falling and thus
 there are 
time dependent boundary condition at the shell:
this  makes a straightforward
Fourier decomposition in time impossible.
However, a possible simplification can be made if the Fourier mode 
is much faster than the falling of shell, i.e. 
the condition $\omg\gg\frac{dz}{dt}$ holds, 
we may consider the shell as quasi-static. In this limit, we can trust
the Fourier mode and study the problem in the usual way. In other
words, we only trust quantities obtained for $\omg\gg\frac{dz}{dt}$.
With this argument,
we can in principle study two-point function in any stage provided the mode
is fast enough.

\subsection{Matching Condition on the Shell}

Israel junction condition in general should be applicable for any
gravity fields. Therefore, one can apply it for
background plus graviton perturbation and from the latter
obtain 
the matching condition for the graviton.

Similar as in Sec.\ref{jc}, we start with the Lanczos equation:
\be
[K_{ij}-g_{ij}K]=-\ka_5^2 S_{ij}=-\ka_5^2 pg_{ij}
\ee
which we cast  into a different form:
\be\label{eq:lanczos2}
[K_{ij}]=\frac{\ka_5^2 p}{3}g_{ij}
\ee
The zeroth order (in $h_{mn}$) Lanczos equation have
already been used above, for calculation of 
 the trajectory of the shell. 
Now we require vanishing of the first order 
terms in Lanczos equation
\be
K_{ij}=n_\alpha(\frac{\del^2x^\alpha}{\del\xi^i\del\xi^j}
+\Gamma^\alpha_{\beta\gamma}\frac{\del x^\beta}{\del\xi^i}\frac{\del x^\gamma}{\del\xi^j})
\ee
with $\alpha=t,z,{\vec x}$ and $i=\tau,{\vec x}$. 
$n_\alpha=(-\frac{\dot z}{z},{\vec 0},\frac{\dot t}{z})$ remains unchanged,
so the variation of $K_{ij}$ comes entirely from Christoffel:
\be\label{var_K}
\dlt K_{ij}=n_\alpha\dlt\Gamma^\alpha_{\beta\gamma}\frac{\del x^\beta}{\del\xi^i}\frac{\del x^\gamma}{\del\xi^j}
\ee

We choose the gauge $h_{\mu z}=0$ and further assume 
$h_{mn}=h_{mn}(t,w,z)$, where
$x\equiv x_1$,$y\equiv x_2$,$w\equiv x_3$. This will not affect the two-point
function because the gravity background is rotationally invariant in $R^3$ 
Calculating the variation of Christoffel to the first order in $h_{mn}$ and 
noting ${\dot z}=0$,${\dot t}=\frac{\sqrt{f+{\dot z}^2}}{f}=\frac{1}{\sqrt{f}}$
(quasi-static limit), 
we find the only non-vanishing components of $\dlt K_{ij}$ is:

\be
&&\dlt K_{xy}=-\frac{\sqrt{f}z}{2}\del_z h_{xy} \no
&&\dlt K_{\tau x}=-\frac{z}{2}\del_z h_{tx} \no
&&\dlt K_{xw}=-\frac{z\sqrt{f}}{2}\del_z h_{xw} \no
&&\dlt K_{\tau\tau}=-\frac{z}{2\sqrt{f}}\del_z h_{tt} \no
&&\dlt K_{\tau w}=-\frac{z}{2}\del_z h_{tw} \no
&&\dlt K_{xx}=-\frac{z\sqrt{f}}{2}\del_z h_{xx} \no
&&\dlt K_{ww}=-\frac{z\sqrt{f}}{2}\del_z h_{ww} \nonumber
\ee
We have omitted some components involving index $y$: those
 can be obtained by
the substitution $x\rightarrow y$ from those listed above.
Plugging to (\ref{eq:lanczos2}), we have:

\be\label{mc_x}
&&\del_z h_{xy}-\sqrt{f}\del_z h_{xy}^f=-\frac{2\ka_5^2 p}{3z}h_{xy} \no
&&h_{xy}=h^f_{xy} \no
&&\del_z h_{tx}-\del_z h_{tx}^f=-\frac{2\ka_5^2 p}{3z}h_{tx} \no
&& h_{tx}=h_{tx}^f{\dot t}=\frac{h_{tx}^f}{\sqrt{f}} \no
&&\del_z h_{xw}-\sqrt{f}h_{xw}^f=-\frac{2\ka_5^2 p}{3z}h_{xw} \no
&&h_{xw}=h_{xw}^f \no
&&\del_z h_{tt}-\frac{1}{\sqrt{f}}\del_z h_{tt}^f=-\frac{2\ka_5^2 p}{3z}h_{tt} \no
&&h_{tt}=\frac{h_{tt}^f}{f} \no
&&\del_z h_{tw}-\del_z h_{tw}^f=-\frac{2\ka_5^2 p}{3z}h_{tw} \no
&&h_{tw}=\frac{h_{tw}^f}{\sqrt{f}} \no
&&\del_z h_{xx}-\sqrt{f}\del_z h_{xx}^f=-\frac{2\ka_5^2 p}{3z}h_{xx} \no
&&h_{xx}=h_{xx}^f \no
&&\del_z h_{ww}-\sqrt{f}\del_z h_{ww}^f=-\frac{2}{\ka_5^2 p}{3z}h_{ww} \no
&&h_{ww}=h_{ww}^f
\ee

We use from here on $h_{mn}$ and $h^f_{mn}$ for metric perturbations
inside and outside the
shell respectively. All the quantities are evaluated on the shell $z=z_m$.
The other identities follow from the continuity of induced
metric across the shell.

Note the jump in time coordinate, we have to do the Fourier transform in a
consistent way\footnote{The convention we use is 
$h_{mn}(t,w)=\int \wh_{mn}(\omg,q)e^{i\omg t-iqw}dtdw$}: 
$\int dt_{out}e^{i\omg t_{out}}=\frac{1}{\sqrt{f}}\int dt_{in}e^{i{\omg/\sqrt{f}}t_{in}}$, 
The indices ``in'' and ``out'' represent quantities inside and outside the
shell. We use $\omg$ as the frequency measured by clock outside the 
shell, the corresponding frequency inside is given by $\frac{\omg}{\sqrt{f}}$. Thus we obtain from (\ref{mc_x}):

\be\label{mc_p}
&&\wh_{xy}^f=\frac{1}{\sqrt{f}}\wh_{xy} \no
&&\del_z\wh_{xy}^f=\frac{1}{f}(\del_z\wh_{xy}+\frac{2\ka_5^2 p}{3z}\wh_{xy}) \no
&&\wh_{tx}^f=\wh_{tx} \no
&&\del_z\wh_{tx}^f=\frac{1}{\sqrt{f}}(\del_z\wh_{tx}+\frac{2\ka_5^2 p}{3z}\wh_{tx}) \no
&&\wh_{xw}^f=\frac{\wh_{xw}}{\sqrt{f}} \no
&&\del_z\wh_{xw}^f=\frac{1}{f}(\del_z\wh_{xw}+\frac{2\ka_5^2 p}{3z}\wh_{xw}) \no
&&\wh_{tt}^f=\sqrt{f}\wh_{tt} \no
&&\del_z\wh_{tt}^f=\del_z\wh_{tt}+\frac{2\ka_5^2 p}{3z}\wh_{tt} \no
&&\wh_{tw}^f=\wh_{tw} \no
&&\del_z\wh_{tw}^f=\frac{1}{\sqrt{f}}(\del_z\wh_{tw}+\frac{2\ka_5^2 p}{3z}\wh_{tw}) \no
&&\wh_{xx}^f=\frac{1}{\sqrt{f}}\wh_{xx} \no
&&\del_z\wh_{xx}^f=\frac{1}{f}(\del_z\wh_{xx}+\frac{2\ka_5^2 p}{3z}\wh_{xx}) \no
&&\wh_{ww}^f=\frac{1}{\sqrt{f}}\wh_{ww} \no
&&\del_z\wh_{ww}^f=\frac{1}{f}(\del_z\wh_{ww}+\frac{2\ka_5^2 p}{3z}\wh_{ww})
\ee

All the quantities are evaluated on the shell $z=z_m$

From here on, we define $u=\frac{z^2}{z_h^2}=z^2(\pi T)^2$ in accordance 
with the literature\cite{scalar}. The axial gauge is just $h_{mu}=0$. The metric 
perturbations can be classified into three channels, according to \cite{kovtun}:
scalar channel: $h_{xy}$
shear channel: $h_{tx}$ and $h_{xw}$
sound channel include $h_{tt}$, $h_{tw}$, $h_{ww}$ and $h_{aa}=h_{xx}+h_{yy}$

The metric perturbations satisfy linearized Einstein equation,
and they are
determined up to residual gauge transformation 
$h_{mn}\rightarrow h_{mn}-\nabla_m\xi_n-\nabla_n\xi_m$
where $\xi_m$ should preserve the axial gauge chosen above.
 Instead of fixing the
gauge, one can look for gauge invariant combination in each channel.
 The behavior
of these gauge invariant objects encodes complete information of retarded 
correlator\cite{kovtun,spectral}.\footnote{there is a subtlety in the
sound channel, which will be elaborated later}

\subsubsection{The scalar channel}
 The gauge invariant object is simply
$\phi_3^f=\wh_{xy}^f$ outside the shell and $\phi_3=\wh_{xy}$ 
inside.
The EOM of $\phi_3^f$ is given by(with $f=1$ corresponding to $\phi_3$)
\be\label{phi3_eom}
\phi_3^f{''}+\frac{1}{u}(3-\frac{2}{f})\phi_3^f{'}+\frac{f-2}{u^2f}\phi_3^f
-\frac{q^2f-\omg^2}{uf^2}\phi_3^f=0
\ee
The prime denote derivative with respect to $u$.
The matching condition between $\phi_3^f$ and $\phi_3$ can be easily obtained
from (\ref{mc_p}):

\be\label{mc_scalar}
&&\phi_3^f=\frac{1}{\sqrt{f}}\phi_3 \no
&&\phi_3^f{'}=\frac{1}{f}(\phi_3'+\frac{\ka_5^2 p}{3u}\phi_3)
\ee

Besides the matching condition (\ref{mc_scalar}), we also need boundary 
condition at
AdS center to uniquely fix the solution of $\phi_3^f(\phi_3)$ in the bulk up to
normalization. The boundary condition we use at AdS center $z=\infty$ is 
infalling wave or regular wave. With these conditions, we are ready to
proceed. Let us start from inside the shell: $\phi_3$ is given by 
the solution to the following equation:

\be\label{inside}
\phi_3''+\frac{1}{u}\phi_3'-\frac{\phi_3}{u^2}+\(\frac{\omg^2}{f_m}-q^2\)\frac{\phi_3}{u}=0
\ee

where $f_m=f(u_m)$. It origins from the jump of frequency across the shell.
The EOM can be solved in terms of cylindrical function:

\be\label{wh}
\phi_3=\left\{\begin{array}{l@{\quad\quad}l}
H_2^{(2)}(2\sqrt{\omg^2/f_m-q^2}\sqrt{u})& \frac{\omg}{\sqrt{f_m}}>\lv q \rv\\
H_2^{(1)}(2\sqrt{\omg^2/f_m-q^2}\sqrt{u})& \frac{\omg}{\sqrt{f_m}}<-\lv q \rv\\
K_2(2\sqrt{q^2-\omg^2/f_m}\sqrt{u})& \lv q \rv>\frac{\omg}{\sqrt{f_m}}
\end{array}
\right.
\ee

While outside the shell, $\phi_3^f$ can be written as a linear combination
of two independent solutions, which we
select to be the infalling wave and the outfalling waves at
the horizon\footnote{Note that it is just a formal basis for
the solution above, 
we don't use it below the shell and there is no horizon
singularity there.}
\be
\phi_3^f=c_+\phi_3^+ +c_-\phi_3^-
\ee

where $\phi_3^{\pm}$ are solutions to (\ref{phi3_eom}). If extrapolated to $u>u_m$, $\phi_3^{\pm}\sim (1-u)^{\pm i\omg/2}$ as $u\rightarrow 1$. (\ref{mc_scalar}) gives:

\be
\frac{c_+}{c_-}=-\frac{\phi_3^-P-\phi_3^-{'}Q}{\phi_3^+P-\phi_3^+{'}Q}\vert_{u=u_m}
\ee

where $P=\frac{1}{f}\(\phi_3'-\frac{\ka_5^2 p}{u}\phi_3\)$,
$Q=\frac{1}{\sqrt{f}}\phi_3$.

We expect to recover the equilibration because, as the shell 
approaches the ``horizon'', the region where geometry deviates
from AdS-BH shrinks exponentially. The ratio $\frac{c_+}{c_-}\rightarrow\infty$.
 All deviations from 
equilibrium should become exponentially small as well,
as we expect from any other small deviations from equilibrium.

We would like to confirm this limit by
our matching/boundary condition. As the shell approaches the ``horizon'',
$u_m\rightarrow 1$,$f_m\rightarrow 0$, we may disregard the third situation in
(\ref{wh}). We want to calculate the ratio $\frac{c_+}{c_-}$ 
to the leading order
in $f_m$ or $(1-u_m)$. The correction to $\phi_3^{\pm}(u)=(1-u)^{\pm i\omg/2}$ is
linear in $f_m$, while the correction from asymptotic expansion of
Hankel function is of $\sqrt{f_m}$. Therefore we may simply use
$\phi_3^{\pm}=(1-u)^{\pm i\omg/2}$.

Let's focus on the first situation $\phi_3=H_2^{(2)}(2\lam\sqrt{u})$, with
$\lam\equiv\sqrt{\frac{\omg^2}{f_m}-q^2}=\frac{\omg}{\sqrt{f_m}}(1+O(f_m))$. The 
correction due to $q^2$ can also be ignored at leading order. Using the
asymptotic expansion of Hankel function, we find the leading order result
cancels exactly in the denominator, while the counterpart survives in the
 numerator. We end up with:

\be\label{asymp_ratio}
\frac{c_+}{c_-}=(1-u_m)^{-i\omg}\frac{-i\omg}{\sqrt{f_m}}\frac{1}{1/8-\ka_5^2 p/6}
\ee

The asymptotic ratio (\ref{asymp_ratio}) tends to infinity universally 
for any REAL $\omg$, correctly recovering the AdS-BH limit.

\subsubsection{The shear and the sound channels}
 The gauge invariant object in the former case is 
$\phi_1^f=q\wh_{tw}^f+\omg\wh_{xw}^f$ outside the shell and 
$\phi_1=q\wh_{tw}+\frac{\omg}{\sqrt{f_m}}\wh_{xw}$ inside. Here
again the frequency inside is scaled by $\frac{1}{\sqrt{f_m}}$. The matching
condition between $\phi_1^f$ and $\phi_1$ from (\ref{mc_p}) turns out
to be the same as the scalar case, up to constant normalization:

\be\label{mc_shear}
&&\phi_1^f=\phi_1 \no
&&\phi_1^f{'}=\frac{1}{\sqrt{f}}(\phi_1'+\frac{\ka_5^2 p}{3u}\phi_1)
\ee

The EOM of $\phi_1^f$ is given by(with $f=1$ corresponding to $\phi_1$):

\be\label{phi1_eom}
&&\phi_1^f{''}+\frac{f(3\omg^2-q^2)-2\omg^2}{uf(\omg^2-q^2f)}\phi_1^f{'} \no
&&+\frac{(\omg^2-q^2f)^2u+f^3q^2+f^2\omg^2-2f\omg^2}{u^2f^2(\omg^2-q^2f)}\phi_1^f=0
\ee

With enough luck, we note $\phi_1$ satisfies the same EOM as $\phi_3$
(\ref{inside}), which combined with
(\ref{mc_shear}) guarantees the same asymptotic ratio (\ref{asymp_ratio}).

Finally we consider the sound channel. The gauge invariant object is

$\phi_2^f=\frac{1}{\sqrt{f_m}}\(2q^2\wh_{tt}^f+4\omg q\wh_{tw}^f+2\omg^2\wh_{ww}^f+(q^2(2-f)-\omg^2)\wh_{aa}^f\)$
outside the shell and 
$\phi_2=2q^2\wh_{tt}+\frac{4\omg q}{\sqrt{f_m}}\wh_{tw}+\frac{2\omg^2}{f_m}\wh_{ww}+(q^2-\frac{\omg^2}{f_m})\wh_{aa}$ inside the shell.

This time we do not seem to have simple matching condition as (\ref{mc_scalar})
and (\ref{mc_shear}). An exception is at $q=0$, in which case, we have:

\be\label{mc_sound}
&&\phi_2^f=\phi_2 \no
&&\phi_2^f{'}=\frac{1}{\sqrt{f}}(\phi_2{'}+\frac{\ka_5^2 p}{3u}\phi_2)
\ee

For case $q\neq 0$, an easy way to avoid general discussion is to take
advantage of the residue gauge. It can be shown by a proper choice of
residue gauge, (\ref{mc_sound}) still holds. The particular gauge choice
of course does not shift the retarded correlator. We include a brief
justification for (\ref{mc_sound}) in Appendix.\ref{gc}

The EOM of $\phi_2^f$ is given by(with $f=1$ corresponding to $\phi_2$):

\be\label{phi2_eom}
&&\phi_2^f{''}-\frac{q^2f^2-8q^2f+4q^2+9\omg^2f-6\omg^2}{uf(q^2f+2q^2-3\omg^2)}\phi_2^f{'}- \no
&&\frac{3\omg^2f^2-6\omg^2f-4q^2\omg^2uf-2q^2\omg^2u+3\omg^4u+2q^4fu-q^2f^3+q^4f^2u+4q^2f^2}{u^2f^2(q^2f+2q^2-3\omg^2)}\phi_2^f=0
\ee

We again find $\phi_2$ satisfies the same EOM as $\phi_3$
(\ref{inside}), which combined with
(\ref{mc_shear}) guarantees the same asymptotic ratio (\ref{asymp_ratio}).

Summarizing the discussion above, we have found the EOM of gauge invariant
objects for three channels:(\ref{phi3_eom}),(\ref{phi1_eom}) and (\ref{phi2_eom}). We also find universal matching condition for all three channels(up to
a constant normalization):

\be\label{mc_all}
&&\phi_a^f=\phi_a \no
&&\phi_a^f{'}=\frac{1}{\sqrt{f}}(\phi_a{'}+\frac{\ka_5^2 p}{3u}\phi_a)
\ee

where $a=1,2,3$ and $\phi_a$ is given by (\ref{wh}).
We may from now on forget about the geometry inside the shell and simply
consider (\ref{mc_all}) as a boundary condition at the shell.

\subsection{Retarded Correlators and their spectral densities}\label{sec_rc}

In this section, we will solve for the gauge invariant objects and extract
the retarded correlator. As in \cite{scalar}, we switch from $\phi_a$ to 
$Z_a=u\phi_a/{(\pi T)^2}$, which couples
to the boundary stress tensor. The equation satisfied by $Z_a$ can be
simply derived from (\ref{phi3_eom}), (\ref{phi1_eom}) and (\ref{phi2_eom}):

\be\label{Z_eom}
&&Z_3{''}-\frac{1+u^2}{uf}Z_3{'}+\frac{\omg^2-q^2f}{uf^2}Z_3=0 \\
&&Z_1{''}-\frac{(\omg^2-q^2f)f-u\omg^2f'}{uf(\omg^2-q^2f)}Z_1{'}+\frac{\omg^2-q^2f}{uf^2}Z_1=0\\
&&Z_2{''}-\frac{3\omg^2(1+u^2)+q^2(2u^2-3u^4-3)}{uf(3\omg^2+q^2(u^2-3))}Z_2{'}+\no
&&\frac{3\omg^4+q^4(3-4u^2+u^4)+q^2(4u^2\omg^2-6\omg^2-4u^3f)}{uf^2(3\omg^2+q^2(u^2-3))}Z_2=0
\ee

Without a simple analytical expression of the ratio between the infalling
and outfalling waves, we study numerically the solutions to (\ref{Z_eom}).
The boundary conditions from (\ref{mc_all}) are given by:

\be\label{Z_bc}
&&Z_a^f=\frac{u}{\sqrt{f}}h \no
&&Z_a^f{'}=\frac{u}{f}h{'}+\(\frac{1}{\sqrt{f}}+\frac{\ka_5^2 p}{3f}\)h 
\ee

where $a=1,2,3$. We have written the boundary conditions for the three
channels in a uniform way. This is achieved by an appropriate scaling
in $Z_a$ and does not affect the two-point functions.

The retarded correlator are obtained according to the prescription
specified in
\cite{spectral}:
\be\label{G_a}
G_a=-\pi^2N_c^2T^4\lim_{u\rightarrow 0}\(\frac{Z_a''}{2Z_a}-h_a\ln(u)\)
\ee
where $h_a=-\frac{1}{2}(q^2-\omg^2)^2$. 

The retarded correlators are extracted from the boundary behavior of 
numerical solutions to (\ref{Z_eom}). The retarded correlators 
correspond to some off-equilibrium state. We compare them with the
counterparts in equilibrium state, which are obtained from numerical
solutions to (\ref{Z_eom}), but with infalling boundary conditions.

In particular, we study the retarded correlator of momentum density,
energy density and transverse stress. They are related to the gauge invariant
correlator by\cite{kovtun}:

\be
&&G_{tx,tx}=\frac{1}{2}\frac{q^2}{\omg^2-q^2}G_1 \\
&&G_{tt,tt}=\frac{2}{3}\frac{q^4}{(\omg^2-q^2)^2}G_2 \\
&&G_{xy,xy}=\frac{1}{2}G_3
\ee

We focus on the spectral density, which is defined by:

\be
\chi_{\mu\nu,\rho\lam}(\omg,q)=-2ImG_{\mu\nu,\rho\lam}(\omg,q)
\ee

\begin{figure}
\includegraphics[width=0.3\textwidth]{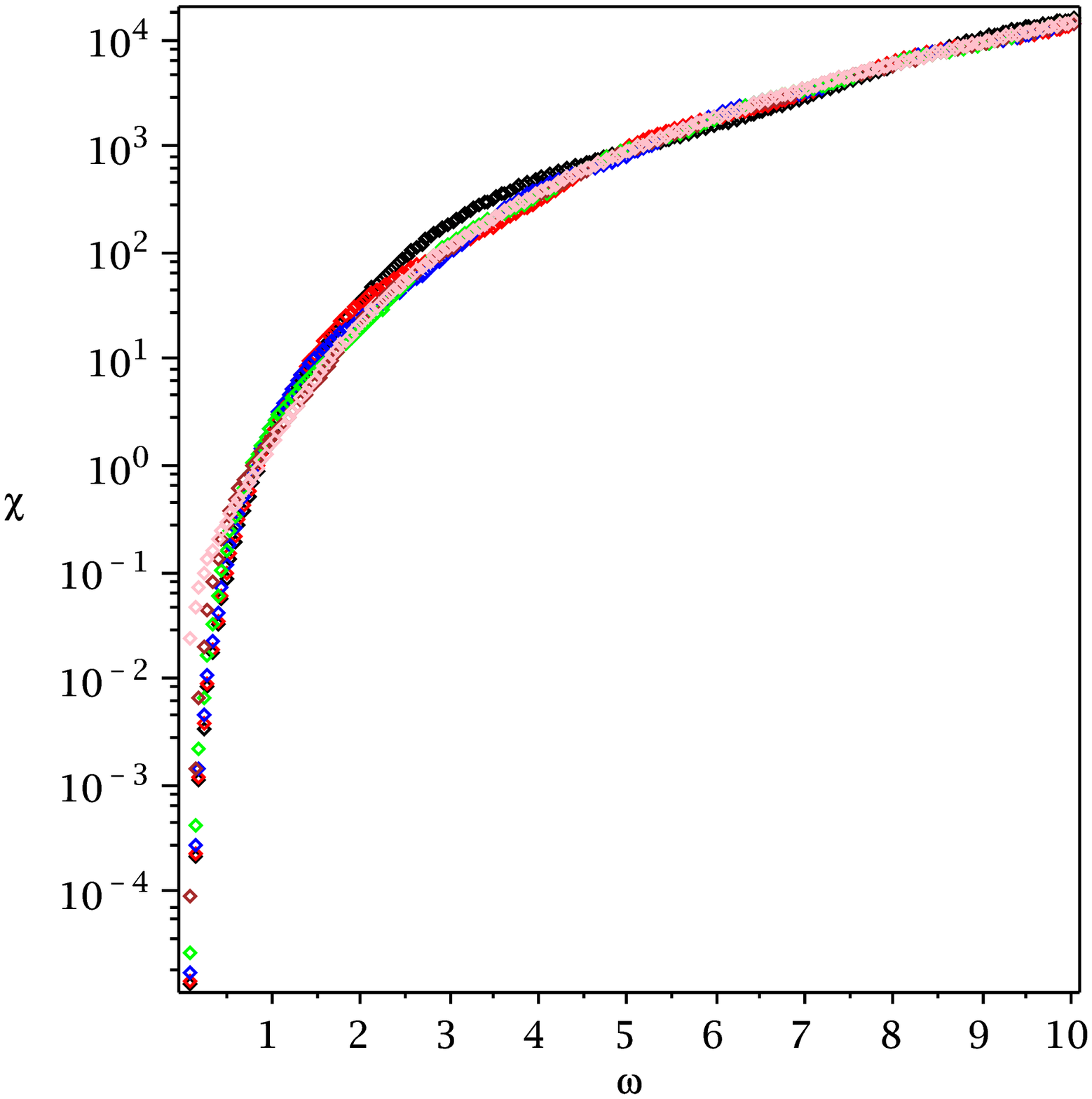}
\includegraphics[width=0.3\textwidth]{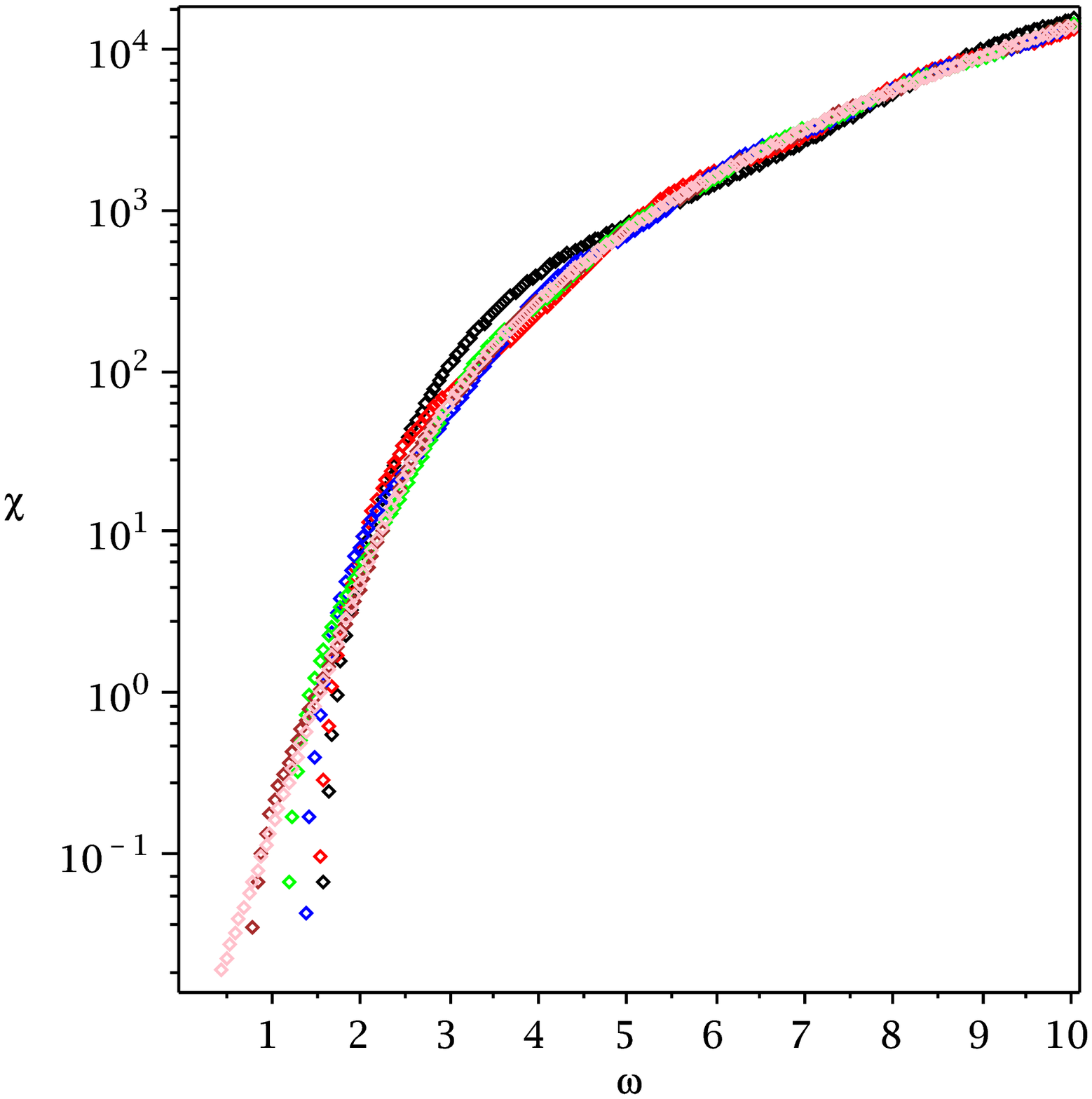}
\includegraphics[width=0.3\textwidth]{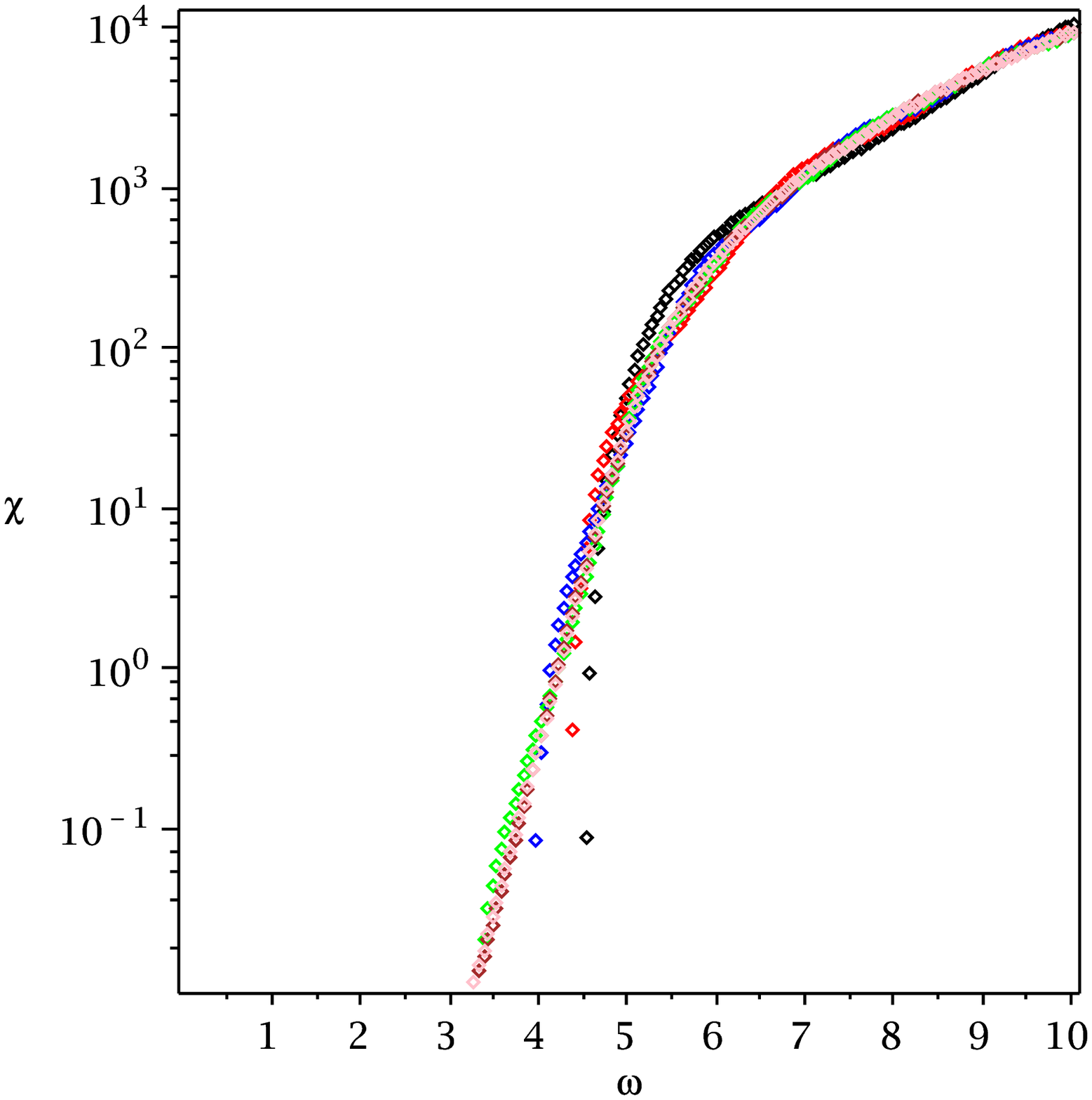}
\caption{\label{scalar_th}(color online)The spectral density of 
transverse stress $\chi_{xy,xy}$ in unit of $\pi^2N_c^2T^4$, at $q=0$ left,
$q=1.5$ middle and $q=4.5$ right. Plotted are spectral densities
at different stages of thermalization: black asterisk($u_m=0.1$), red box($u_m=0.3$),
blue circle($u_m=0.5$), green cross($u_m=0.7$), brown diamond($u_m=0.9$). The thermal 
spectral density is also included(pink point) for comparison. The parameter
we will keep using from here on is $\ka_5^2 p=1$}
\end{figure}

\begin{figure}
\includegraphics[width=0.3\textwidth]{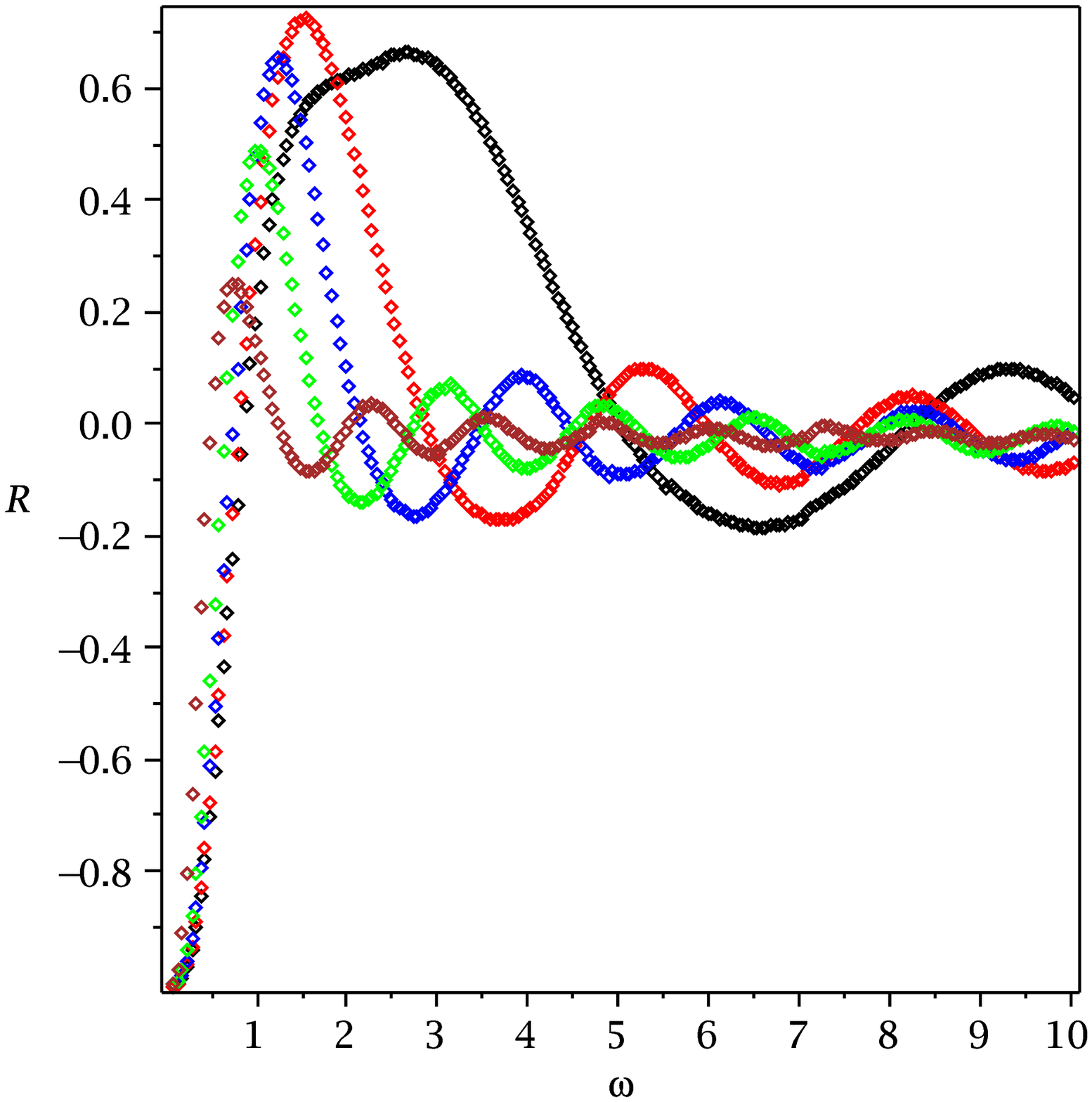}
\includegraphics[width=0.3\textwidth]{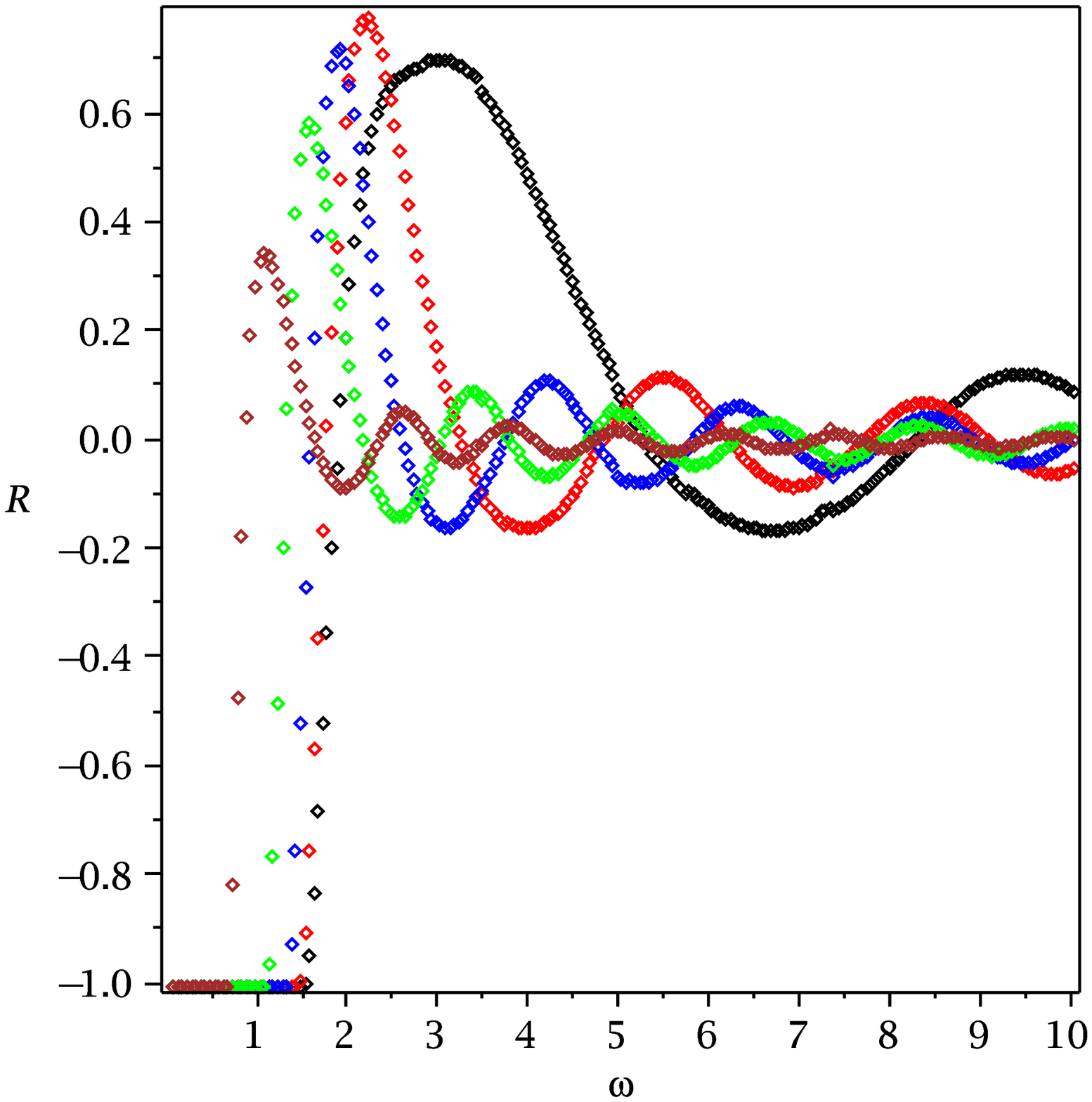}
\includegraphics[width=0.3\textwidth]{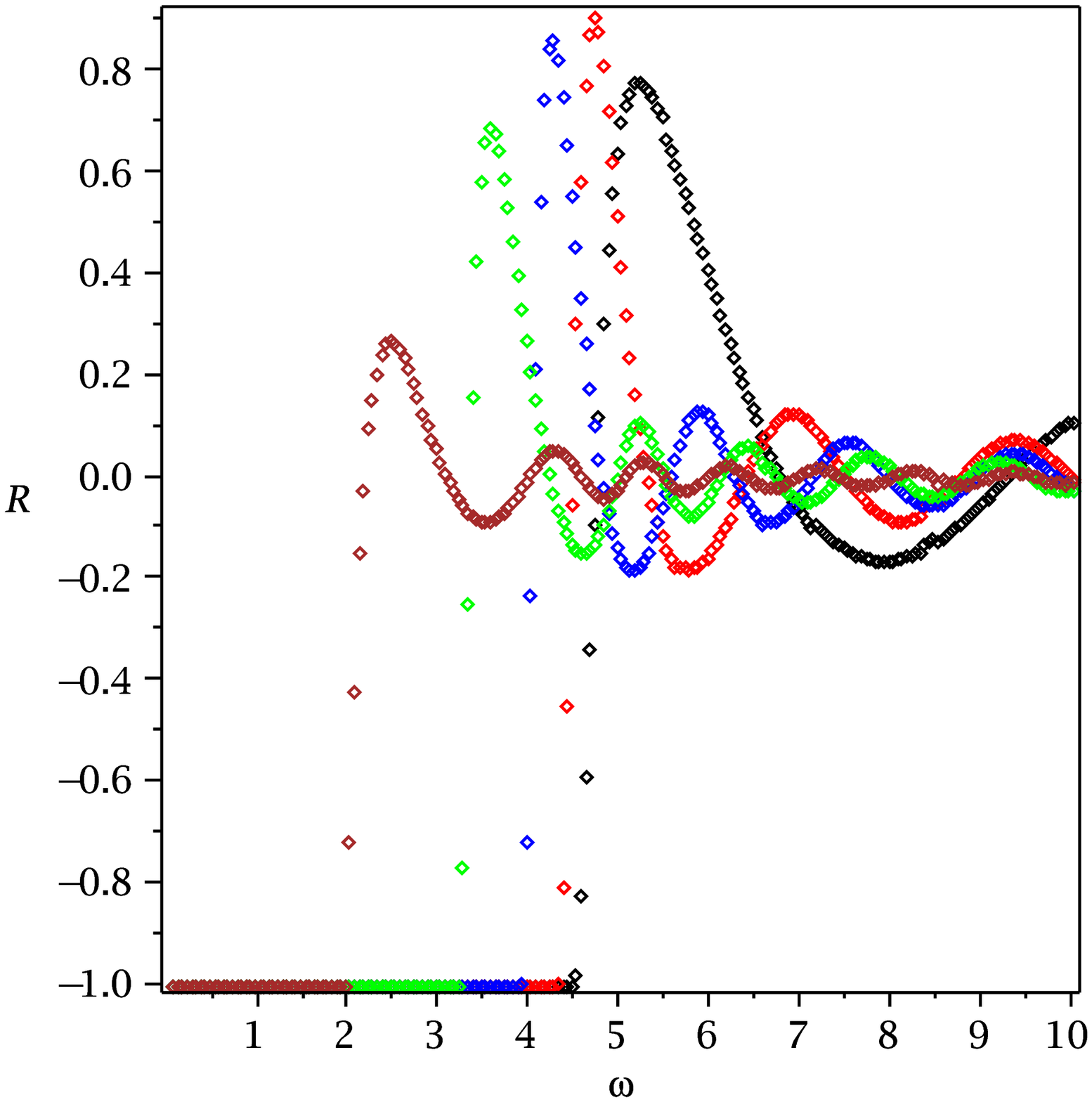}
\caption{\label{scalar_q}(color online)The relative deviation $R$
at $q=0$ left, $q=1.5$ middle and $q=4.5$ right. Different stages of 
thermalization are indicated by: black asterisk($u_m=0.1$), red box($u_m=0.3$),
blue circle($u_m=0.5$), green cross($u_m=0.7$), brown diamond($u_m=0.9$). As $u_m$ approaches 1,
i.e. the medium evolves to equilibrium, the oscillation decreases in
amplitude and increases in frequency, thus the spectral density relaxes to
thermal one}
\end{figure}

The spectral densities of the transverse stress are plotted 
at $q=0$, $q=1.5$ and $q=4.5$ in Fig.\ref{scalar_th}. Each plot includes five
values of $u_m$, corresponding to different stage of thermalization.
The thermal spectral density of
the transverse stress $\chi^{th}_{xy,xy}$ is also 
included for comparison. All the non-thermal spectral densities can be 
viewed as some oscillation on top of their thermal counterpart. 
The first period of
oscillation occurs near $\omg=q$\footnote{Similar behavior is also
observed in thermal spectral density\cite{spectral,hartnoll}}. The oscillation
damps in amplitude and grows in frequency as the 
medium thermalizes\footnote{Here the
frequency refers to the oscillation in spectral density. It is understood
as the reciprocal of $\omg$}, i.e.
$u_m$ approaches $1$ from below. This effect is clearly illustrated in 
Fig.\ref{scalar_q}, where we plot the 
relative deviation
$R\equiv\frac{\chi_{xy,xy}-\chi^{th}_{xy,xy}}{\chi^{th}_{xy,xy}}$.

\begin{figure}
\includegraphics[width=0.4\textwidth]{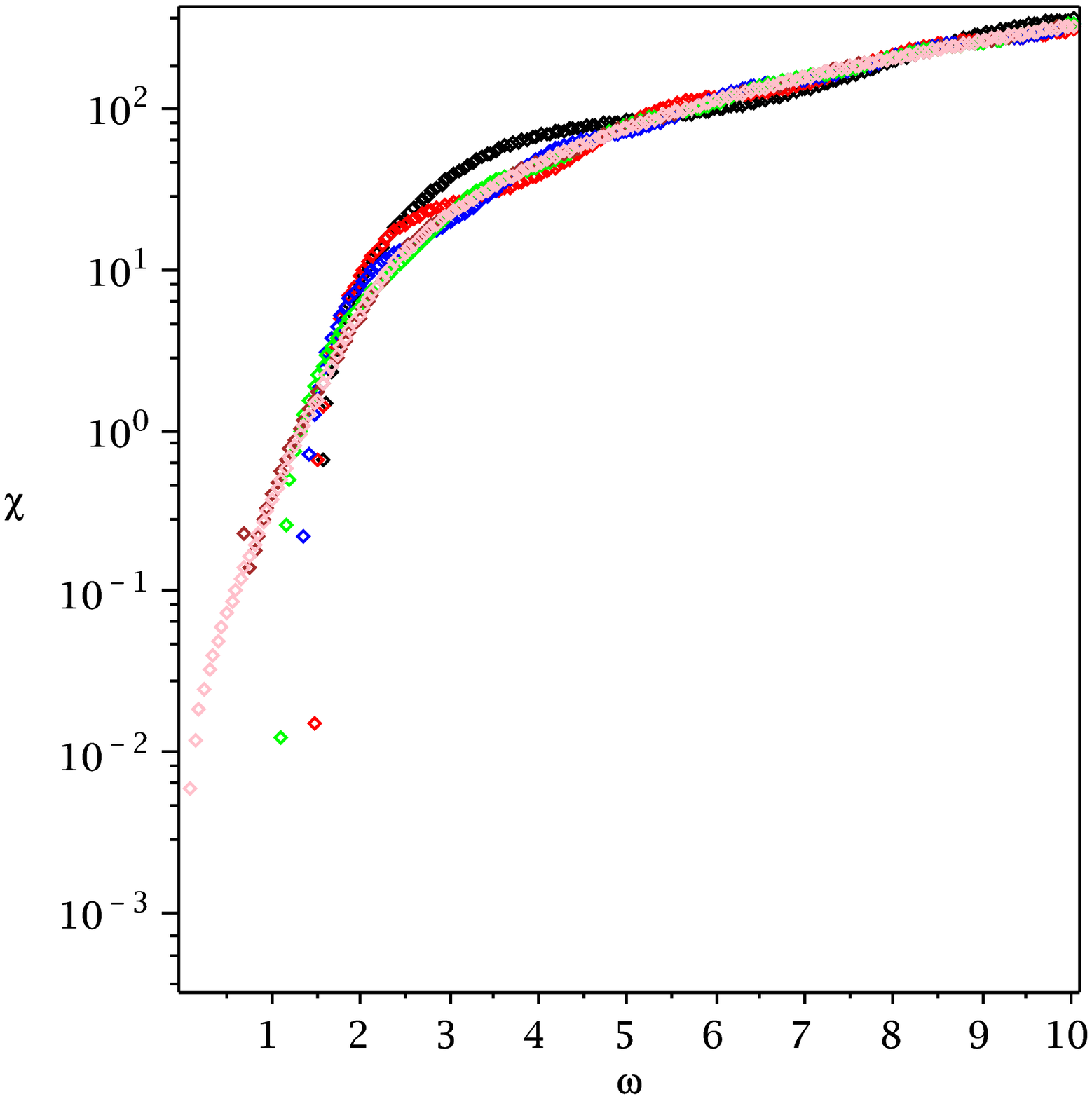}
\includegraphics[width=0.4\textwidth]{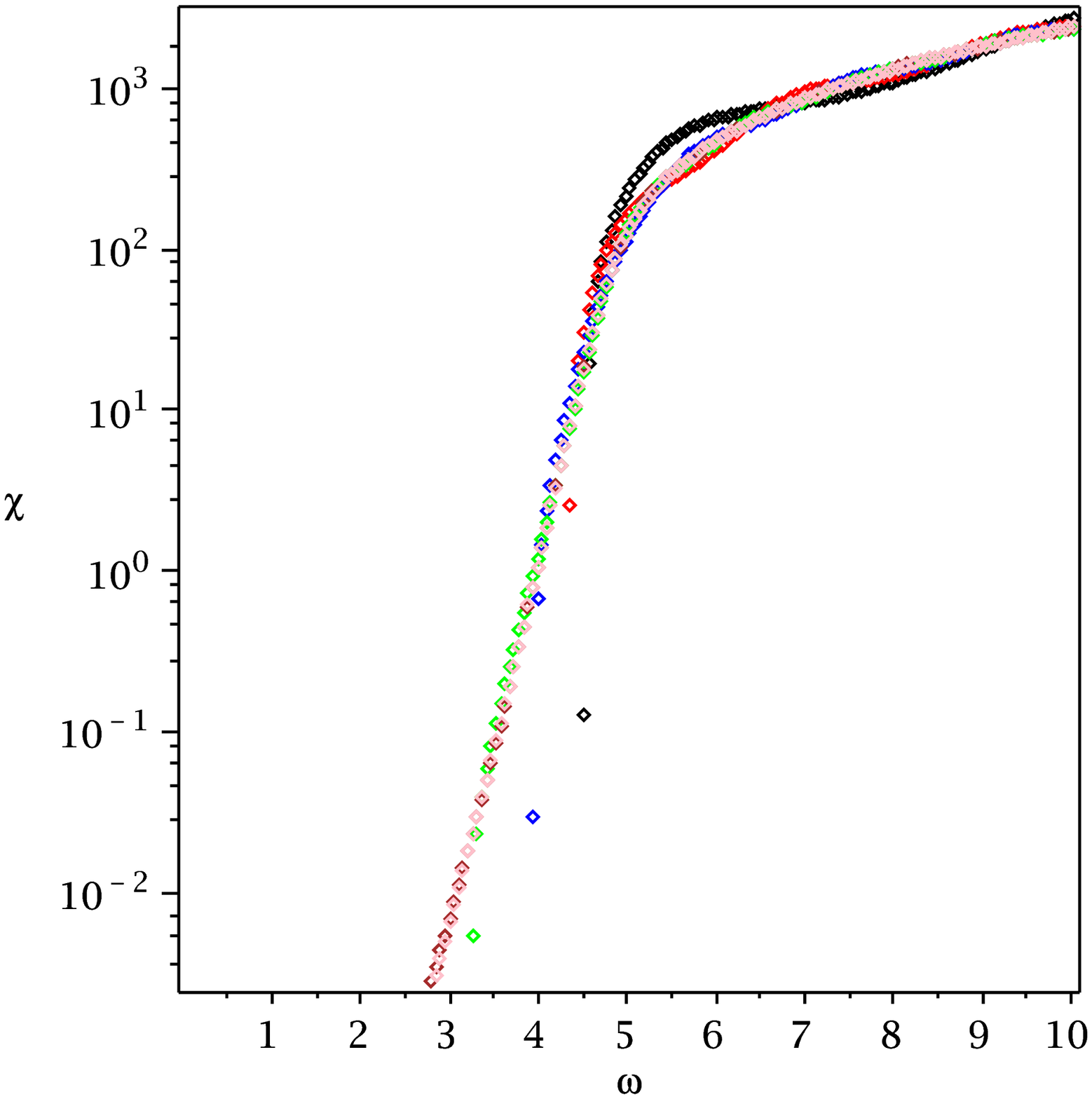}
\caption{\label{shear_th}(color online)The spectral density of 
momentum density $\chi_{tx,tx}$  in unit of $\pi^2N_c^2T^4$ at $q=1.5$ left and
$q=4.5$ right. Plotted are spectral densities
at different stages of thermalization: black asterisk($u_m=0.1$), red box($u_m=0.3$),
blue circle($u_m=0.5$), green cross($u_m=0.7$), brown diamond($u_m=0.9$). The thermal 
spectral density is also included(pink point) for comparison}
\end{figure}

Parallel to the case of transverse stress, we also plot the spectral density
of momentum density, at $q=1.5$ and $q=4.5$ in Fig.\ref{shear_th}
(The spectral density of momentum density at $q=0$ 
vanishes identically). Each plot include five
values of $u_m$, corresponding to different stage of thermalization.

\begin{figure}
\includegraphics[width=0.4\textwidth]{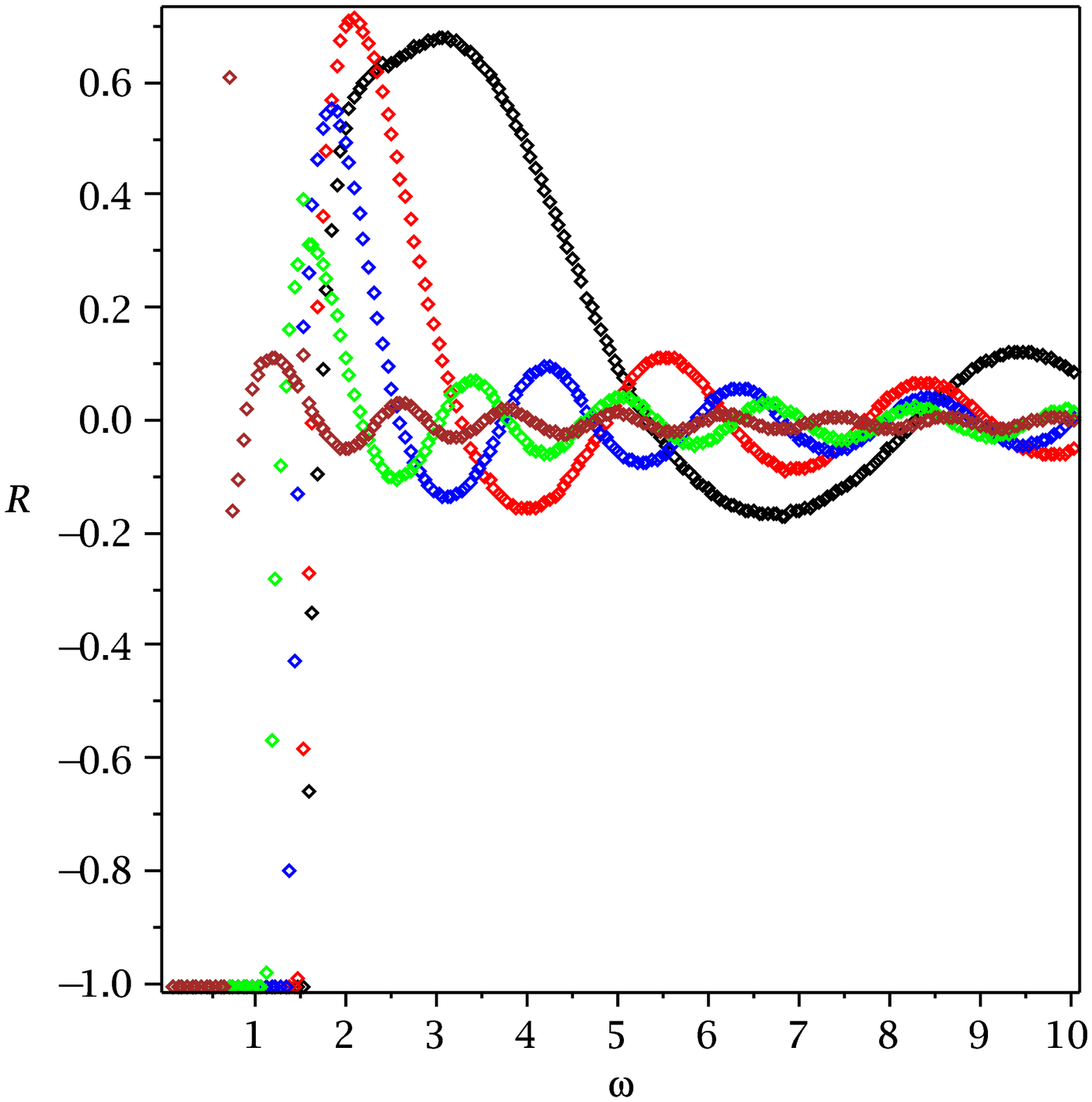}
\includegraphics[width=0.4\textwidth]{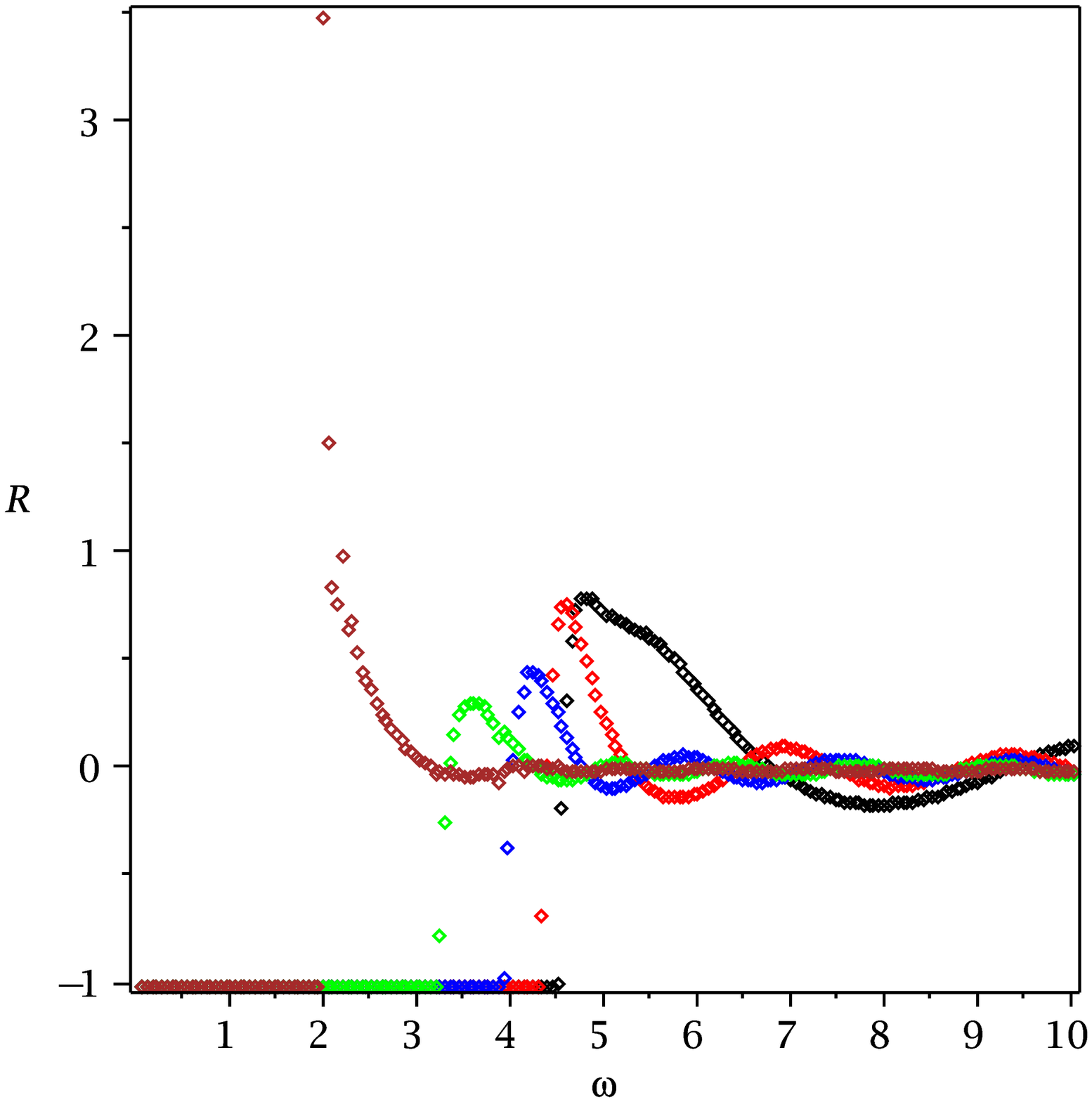}
\caption{\label{shear_q}(color online)The relative deviation $R$
at $q=1.5$ left and $q=4.5$ right. Different stages of thermalization
are indicated by: black asterisk($u_m=0.1$), red box($u_m=0.3$),
blue circle($u_m=0.5$), green cross($u_m=0.7$), brown diamond($u_m=0.9$). As $u_m$ approaches 1,
i.e. the medium evolves to equilibrium, the oscillation decreases in
amplitude and increases in frequency, thus the spectral density relaxes to
thermal one}
\end{figure}

The relative deviation 
$R\equiv\frac{\chi_{tx,tx}-\chi^{th}_{tx,tx}}{\chi^{th}_{tx,tx}}$
is plotted in Fig.\ref{shear_q}.
We again observe the damping of amplitude and
growing of frequency as the medium thermalizes.

\begin{figure}
\includegraphics[width=0.4\textwidth]{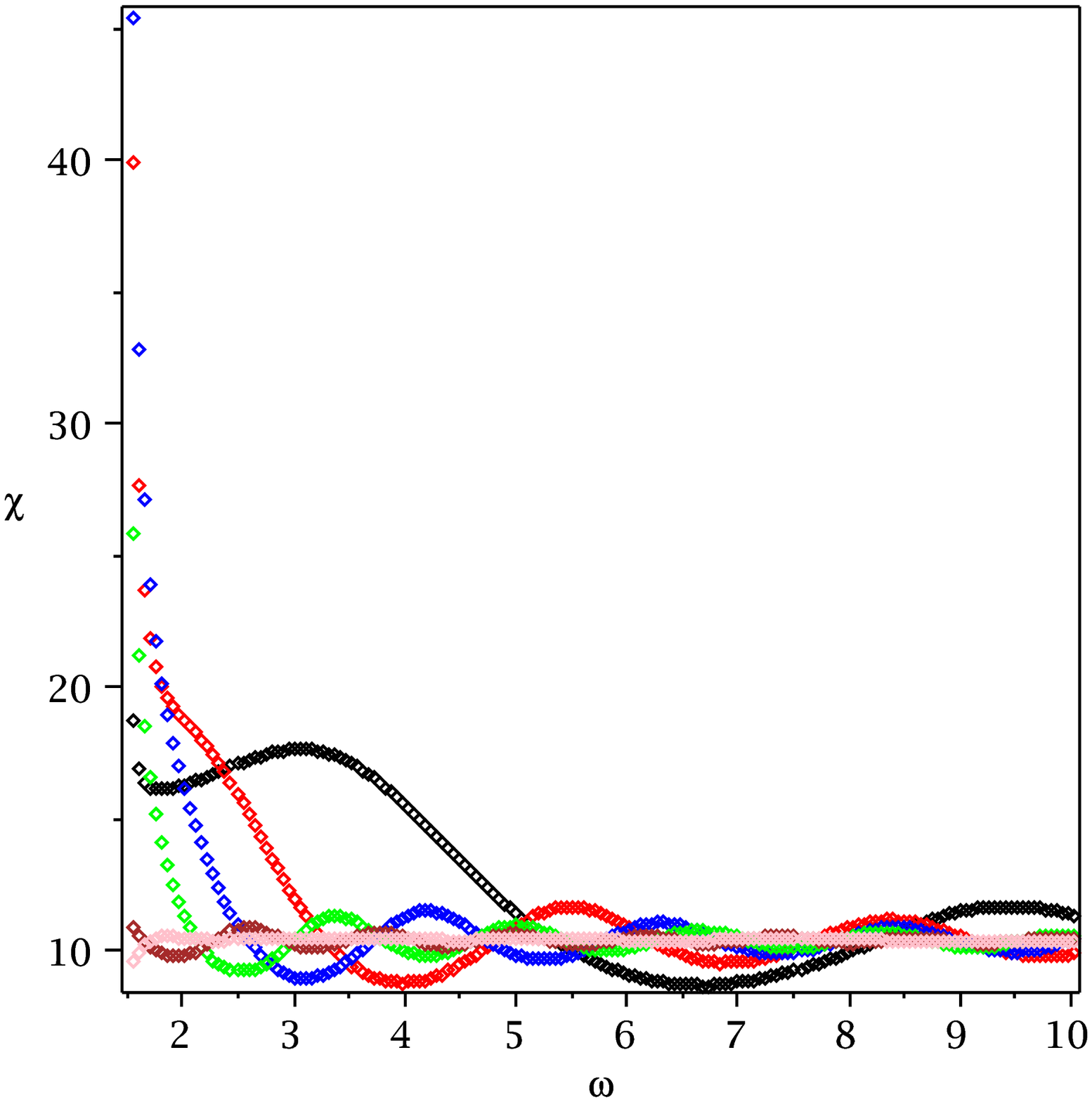}
\includegraphics[width=0.4\textwidth]{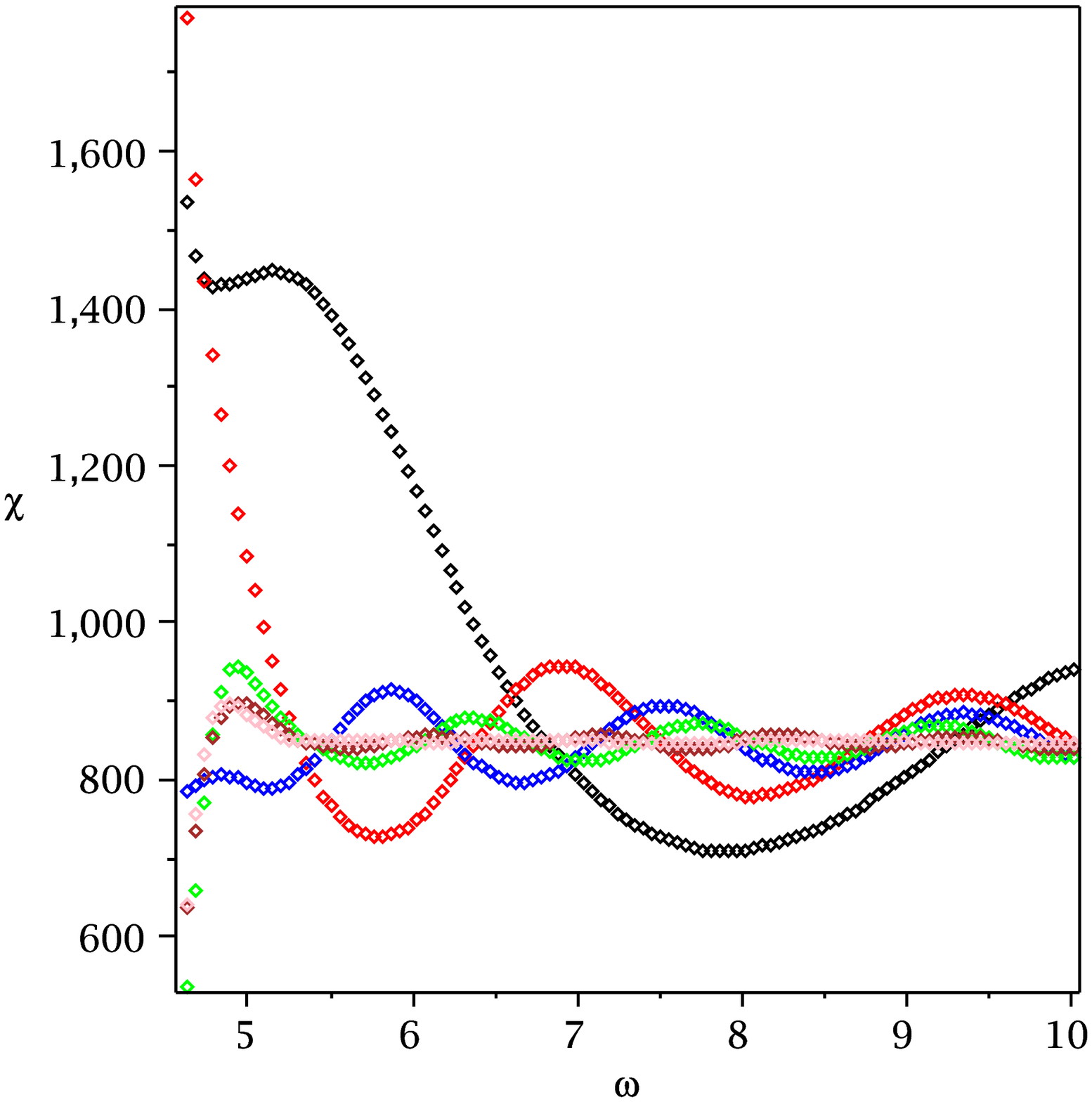}
\caption{\label{sound_th}(color online)The spectral density of 
energy density $\chi_{tt,tt}$ in unit of $\pi^2N_c^2T^4$, at $q=1.5$ left and
$q=4.5$ right. Plotted are spectral densities
at different stages of thermalization: black asterisk($u_m=0.1$), red box($u_m=0.3$),
blue circle($u_m=0.5$), green cross($u_m=0.7$), brown diamond($u_m=0.9$). The thermal 
spectral density is also included(pink point) for comparison}
\end{figure}

\begin{figure}
\includegraphics[width=0.4\textwidth]{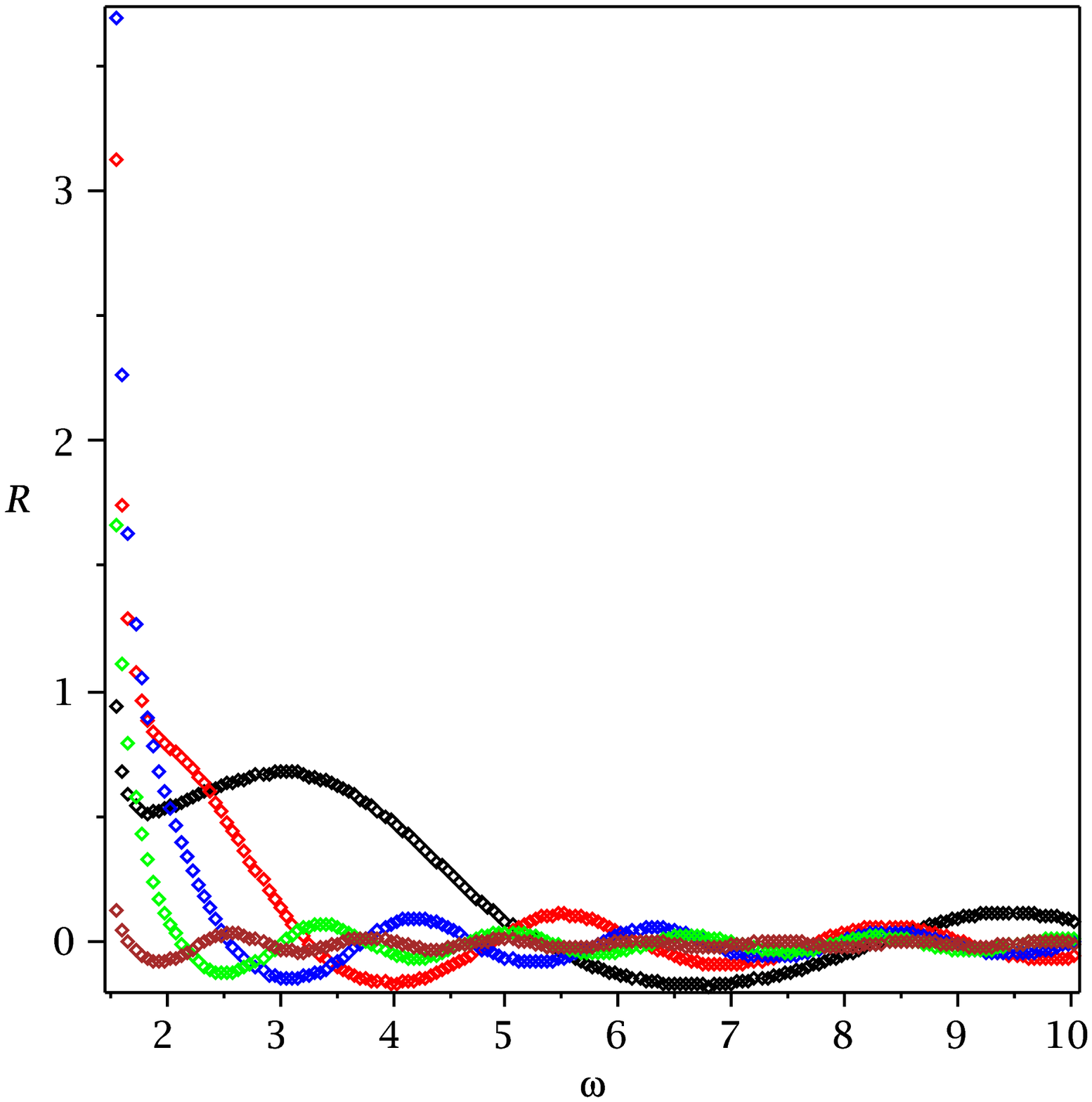}
\includegraphics[width=0.4\textwidth]{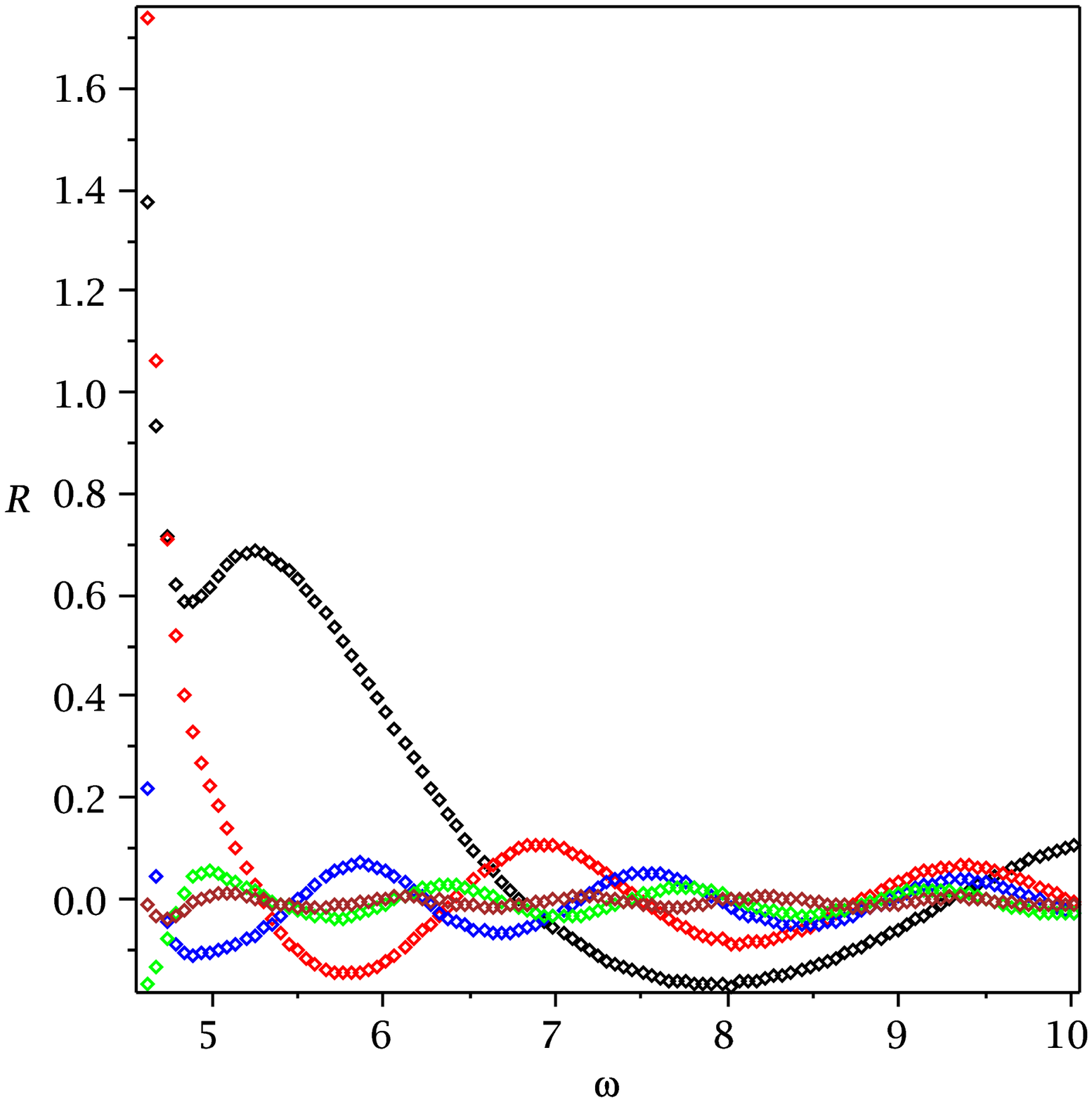}
\caption{\label{sound_q}(color online)The relative deviation $R$
at $q=1.5$ left and $q=4.5$ right. Different stages of thermalization
are indicated by: black asterisk($u_m=0.1$), red box($u_m=0.3$),
blue circle($u_m=0.5$), green cross($u_m=0.7$), brown diamond($u_m=0.9$). As $u_m$ approaches 1,
i.e. the medium evolves to equilibrium, the oscillation decreases in
amplitude and increases in frequency, thus the spectral density relaxes to
thermal one}
\end{figure}

The spectral density of energy density, at $q=1.5$ and $q=4.5$
are plotted in Fig.\ref{sound_th}.
(The spectral density of energy density again vanishes). 
Each plot includes five
values of $u_m$, corresponding to different stage of thermalization. 
The relative deviation 
$R\equiv\frac{\chi_{tt,tt}-\chi^{th}_{tt,tt}}{\chi^{th}_{tt,tt}}$ is plotted 
in Fig.\ref{sound_q}. We find a very sharp peak
in the first period of oscillation, which is removed from the 
final plot for a better comparison.
We again confirm the non-thermal spectral density relaxes to thermal one
in the qualitatively the same way as described in the previous cases.

In order to explain the observed phenomenon, we would like to obtain
some analytical formula for the spectral density. This is
possible in the final freezing stage, where there is a simple
asymptotic ratio (\ref{asymp_ratio}). To the leading order in $1-u_m$, 
the boundary condition as $u\rightarrow 1$ is:

\be
Z_a=c_+Z_a^++c_-Z_a^-\rightarrow c_+(1-u)^{\frac{i\omg}{2}}+c_-(1-u)^{\frac{-i\omg}{2}}
\ee

For the purpose of illuminating the problem, it is enough to focus on
$Z_3$, the EOM of which has the simplest form. Its EOM (\ref{Z_eom}) 
is solvable in terms of Heun function. However
the property of Heun function is not fully understood.\footnote{see \cite{nunez} for a discussion}. 
We have to use some approximation method,
and in the regime of large $\omg$ the WKB treatment is appropriate 
. Following \cite{teaney,minkowski},
we obtain the expression of $Z_3$ near the boundary up to normalization(see
Appendix.\ref{wkb} for details of the treatment):

\be
Z_3=\left\{\begin{array}{l@{\quad}l}
\frac{u}{\sqrt{1-u^2}}\(H_2^{(2)}(2\sqrt{\omg^2-q^2}\sqrt{u})
+\frac{ic_-}{c_+}e^{-2i\omg a_0}H_2^{(1)}(2\sqrt{\omg^2-q^2}\sqrt{u})\)& \omg>\lv q\rv\\
\frac{u}{\sqrt{1-u^2}}\(H_2^{(1)}(2\sqrt{\omg^2-q^2}\sqrt{u})-\frac{ic_-}{c_+}e^{-2i\omg a_0}H_2^{(2)}(2\sqrt{\omg^2-q^2}\sqrt{u})\)& \omg<-\lv q\rv\\
\frac{u}{\sqrt{1-u^2}}\(1-\frac{ic_-}{c_+}e^{-2i\omg b_0}\)\frac{e^{\omg c_0}}{\pi}K_2(2\sqrt{q^2-\omg^2}\sqrt{u})& \\
-\frac{iu}{2\sqrt{1-u^2}}\(1+\frac{ic_-}{c_+}e^{-2i\omg b_0}\)e^{-\omg c_0}I_2(2\sqrt{q^2-\omg^2}\sqrt{u})& 0<\omg<\lv q\rv\\
\frac{u}{\sqrt{1-u^2}}\(1+\frac{ic_-}{c_+}e^{-2i\omg b_0}\)\frac{e^{\omg c_0}}{\pi}K_2(2\sqrt{q^2-\omg^2}\sqrt{u})& \\
+\frac{iu}{2\sqrt{1-u^2}}\(1-\frac{ic_-}{c_+}e^{-2i\omg b_0}\)e^{-\omg c_0}I_2(2\sqrt{q^2-\omg^2}\sqrt{u})& 0>\omg>-\lv q\rv
\end{array}
\right.
\ee

where the constants $a_0,b_0,c_0$ are defined by:

\be
&&lim_{u\rightarrow 1}\int_0^u du \sqrt{\frac{1-s^2(1-u^2)}{u(1-u^2)^2}}=a_0-\frac{1}{2}ln(1-u) \no
&&lim_{u\rightarrow 1}\int_{u_0}^u du \sqrt{\frac{1-s^2(1-u^2)}{u(1-u^2)^2}}=b_0-\frac{1}{2}ln(1-u) \no
&&\int_0^{u_0} du \sqrt{\frac{s^2(1-u^2)-1}{u(1-u^2)^2}}=c_0 \no
&&s=\lv\frac{q}{\omg}\rv \no
&&u_0=\frac{\sqrt{s^2-1}}{s}
\ee

The retarded correlator is calculated from the prescription (\ref{G_a}).
We have dropped an additional contact term: $\pi^2N_c^2T^4(1/2-(\omg^2-q^2)^2\gamma)$.

\be\label{g_wkb}
G^R=\left\{\begin{array}{l@{\quad}l}
\frac{\pi^2 N_c^2T^4(\omg^2-q^2)^2}{4}(ln(\omg^2-q^2)+isgn(\omg)\pi g(sgn(\omg),a_0))& s<1\\
\frac{\pi^2 N_c^2T^4(q^2-\omg^2)^2}{4}(ln(q^2-\omg^2)+isgn(\omg)\pi e^{-2\lv\omg\rv c_0} g(sgn(\omg),b_0))& s>1
\end{array}
\right.
\ee

where we have defined $g(\epsilon,c)=\frac{1+i\epsilon e^{-2i\omg c}c_-/c_+}{1-i\epsilon e^{-2i\omg c}c_-/c_+}$.

In the large $\omg$ limit, the non-logarithmic term in the second case is 
exponentially suppressed. Therefore the lowest order result in $c_-/c_+$
agrees with the zero temperature one.
\cite{minkowski}\footnote{the imaginary part has an opposite sign, which
is due to a different convention in Fourier transform}.  The higher order 
correction is due to the emergence of wave from the ``horizon''. 
The phase difference between infalling and outfalling waves gives rise to 
the oscillating behavior in $\omg$. We have also calculated (\ref{g_wkb}),
with $c_-/c_+$ numerically obtained and restored the contact term. The result
show very good agreement with retarded correlator obtained in Sec.\ref{sec_rc}
in region of large $\omg$. 

The physical interpretation of (\ref{g_wkb}) is most clear at vanishing 
momentum, $s=0$. The n-th order correction appears as 

\be
\dlt^n G^R=\frac{\pi^2 N_c^2T^4\omg^4}{4}2i\pi sgn(\omg^{n+1})\(\frac{ic_-}{c_+}e^{-2i\omg a_0}\)^n 
\ee

Combined with the ratio of $\frac{c_-}{c_+}$ (\ref{asymp_ratio}), the n-th
order correction can be written as 

\be\label{correction}
\dlt^n G^R\sim\omg^{4-n} e^{i\omg(ln(1-u_m)-2a_0)n} 
\ee

Recall the WKB solution of the incoming wave at $s=0$:

\be
&&\psi=S'^{-1/2}e^{-i\omg\int_0^u S'du+i\omg t} \no
&&lim_{u\rightarrow1}\int_0^u S'du=-\frac{1}{2}ln(1-u)+a_0
\ee
Taking into account the time factor $e^{i\omg t}$, we see $-(ln(1-u_m)-2a_0)$ 
is just the time for the wave to travel back-and-forth in WKB potential.
We define the echo time 
\be t_{echo}=(-ln(1-u_m)+2a_0) \ee
 in which the wave
makes a roundtrip. 
The n'th order correction
to the two-point function (spectral density) has an echo time of $n*t_{echo}$,
with a suppressed amplitude, obviously the $n$-th echo.

 This resembles the usual echo phenomenon with a sound
 reflected by a wall. Furthermore, the oscillations
in spectral density as a function of frequency result,
after the Fourier transform, in a $peak$ as a function of time
at the echo time. The peak is a result of many harmonics
coherently added together: while smooth equilibrium spectral densities
correspond to (thermally occupied) field harmonics which
are completely incoherent to each other. We thus interpret
``echo'' as a partial re-appearance of coherence which was
present in the original ``color glass'' fields at the collision
moment.
 As the shell keeps falling toward the ``horizon'',
$u_m\rightarrow 1$, the echo time tends to infinity and the medium 
loses all the coherence.

Although we derived it in near-thermal position
of the shell, the echo phenomenon by itself is not restricted
to the final 
freezing stage but exists throughout 
the whole process: its manifestation is in the spectral density 
of Fig.\ref{scalar_q}, Fig.\ref{shear_q} and Fig.\ref{sound_q}.
Looking for ``echo'' in dynamical (time dependent)
shell is perhaps worth addressing in later work.

\subsection{Can the bounary observer
 see what happens below the shell? }

In the rest of the paper we have chosen the  boundary condition 
to correspond to the infalling wave
 at $u$-infinity (the AdS center), which leads to the
solution (\ref{wh}) and its consequences discussed above.
 This particular choice is natural in the standard setting, when
all the probes (the source and the sink) 
 sending gravitational wave $from$ the AdS boundary $u=0$.

Now we switch to another issue: the gravitational wave emerging
 from inside, below the shell, $u>u_m$. The motivation for studying
this case is as follows. All stationary black hole metrics
are such that no signal from the inside the horizon can
propagate outside it: in particular, the geodesics do not
cross horizon. But in the falling shell case we consider,
 the metric coincides with black hole one only above the shell, while
 inside it is the $AdS_5$ solution $without$ the horizon, since
 gravity of the shell produces no effect inside it. 
The question then is, can an observer on the boundary see
what happens inside the shell? 

Thus a wave sent from below would propagate all the way till the
 shell without any problems, and scatter on it. So we are   
looking now for a solution, which
in the region inside the shell contains both outfalling and infalling
waves, while  outside  ($u<u_m$) it has only outfalling wave
propagating toward the boundary.
At the shell
the matching condition is again given by (\ref{mc_p}), 
and the precise relation 
between outfalling and infalling waves 
can be worked out parallel to what we did above.
As always, a generic graviton is split into three channels,
with the same eqns.

The gravitational wave outside the shell contains
only the outfalling component.
 This is a solution in thermal-AdS metric which,
if extrapolated to (non-existing) region $u_m<u<1$ would be
originating from the horizon. Thus  at $u=1$ it is
$\phi_a^f\sim (1-u)^{-i\omg/2}$, it is understood as the 
 behavior since $u<u_m$. The wave inside the shell can be
written as:
\be
\phi_a=c_{in}\phi_{in}+c_{out}\phi_{out}
\ee
With similar argument as
before, we can approximate $\phi_a^f=(1-u)^{-i\omg/2}$. Matching
solutions at both
sides $\phi_a^f$ and $\phi_a$ according to
Israel condition (\ref{mc_all}), we  obtain
the asymptotic reflection ratio $\frac{c_{in}}{c_{out}}$
\be
\frac{c_{in}}{c_{out}}=\frac{(1/8-\ka_5^2 p/6)\sqrt{1-u_m^2}}{\omg}
exp\(4i\omg\sqrt{\frac {u_m}{1-u_m^2}} \)
\ee
which, at the shell approaching the horizon
 $u_m\rightarrow 1$ tends to zero.
 This means as the shell approaches 
the horizon, the portion of the wave reflected by the shell disappears,
and thus all the wave emerging from below
the shell is transmitted! 

(To convince the reader that this conclusion is correct,
here is another argument. As we have shown in the beginning of the
paper, as the shell approach horizon one recovers the
 solution for the AdS-BH background (without any shell),
with only the
infalling waves on either side of the horizon without reflection
(black horizon). 
The  solution with only outfalling
waves we are now describing is its
 complex conjugate.)

The conclusion that even at late time our collapsing shell
is not that black as a horizon, as
 signals from inside the shell can be seen,
look surprising and worrisome at first.
 Note however that these signals are both
strongly red-shifted and exponentially delayed by
the warping factor $\sqrt{f(u_m)}$. As $u_m\rightarrow 1$, both
effects become infinitely strong, in practical sense
precluding the boundary observer from seeing
what happens below the shell.

\section{Conclusion and Outlook}\label{conclude}
 Continuing the line of research started by our papers I and II,
we have built a gravitational collapse scenario, 
describing equilibration of conformal strongly coupled plasma
in the AdS/CFT setting. Using a simplified geometry we approximated
falling collision debris by a single flat shell or membrane,
falling from its initial position (given by the saturation scale) to 
its horizon (given by the equilibrium temperature). The setting itself
provides inequality between the two scales, satisfied at RHIC.

 The main simplification of the paper is that  
this shell is $flat$ - independent on our world 3 spatial 
coordinates. Therefore the overall solution of Einstein eqns
reduces to two separate regions with well known $static$
AdS-BH and AdS metrics. The falling of
the shell is time dependent, its equation of
motion is determined by the Israel
junction condition, which we solve and analyzed. 
We also determined how final temperature (horizon position)
depends on initial scale and shell tension.

  In statistical mechanics it is known that
fluctuations at different scales are independent, and that
long-range fluctuations (IR) need more time to
be equilibrated than the short-range (UV) ones. 
  With the AdS/CFT setting and our geometric simplification,
this fact is taken to its perfect form. It is
 connected to the well known gravitational phenomenon:
gravity of a sphere is independent on its size outside it,
and is completely absent inside. In the equilibrating gauge theory
it means that mean quantities at all scales above some $z<z_m(t)$
(corresponding to sliding position of the shell)  are $exactly$
as in equilibrium, while those below it are $absolutely$ unaffected,
being the
same as in vacuum without any matter.

This is  $quasiequilibrium$ in the title. More specifically
it means the following.
 In this geometry a ``single point 
observer'' --
who is only able to measure the {\em average density and pressure}
-- would be
driven to the conclusion that the matter is instantaneously  
equilibrated at $all$ times. However more sophisticated  ``two point 
observer'' who is able to study correlation functions
of stress tensors 
would  be able to observe deviations 
from the thermal case. We computed them explicitly,
calculating a number of spectral densities at various positions
of the shell, corresponding (in quasi-static approximation)
to different stages of equilibration.
%

 The equilibration
process
 can roughly be divided into three stages: the initial acceleration,
intermediate relativistic falling and final near-horizon  freezing.
By studying the graviton probes corresponding
to three different combinations of $h_{mn}$,
 we calculate the retarded correlators
of $T_{mn}$ on the AdS boundary, The matching condition of $h_{mn}$ on the shell is given by
a variation of Israel junction condition. In the quasi-static limit,
we study the retarded correlator of all graviton probes. We have shown
that the collapsing geometry correctly
 reproduces the AdS-BH limit: as the
shell approaches the ``horizon'' the ratio of the infalling wave
and outfalling wave tends to infinity. 
We further confirmed this by numerically
study the spectral density for transverse stress, momentum density and 
energy density, which allows us to see deviations
between geometry with shell and equilibrated one (AdS BH).

We find that 
 the main deviation between the  non-thermal and equilibrium
spectral densities are
 some oscillations.
As the time goes on and the shell is at the position closer
and closer to the
horizon,
 these oscillations become
exponentially smaller in amplitude and higher in
frequency, eventually disappearing in the equilibrium state.
In this sense we get numerical control
over the relaxation process.
In the final freezing stage, when the membrane is close
to the horizon, and in large $\omg$ regime, 
we even find analytical expressions for these deviations using the
WKB method. 

Oscillations in spectral densities as a function of frequency
are further explained by the
 ``echo'' effect, producing peaks at certain  ``echo'' times
in the response functions. We expect at those times
partial restoration of field coherence which was present
at the initial time of the collision in system's wave function.  
 For the near equilibrated medium we have
the echo time analytically, from the WKB solution.
 The echo time tends to infinity as the medium thermalizes.

The echo phenomenon arises from the phase 
coherence between infalling and 
outfalling waves in the bulk at certain times: we expect it
to exist in all
gauge theory with a gravity dual. It is also interesting to extend the
current study to scalar probe and vector probe. One
can further study the effect of echo on electromagnetic
($e^+e^-$) spectral densities related to production spectra
\cite{photon} which can be observable in collisions.

An attentive reader will notice that apart of small 
discussion of upward moving waves, we
we have not yet
addressed the dynamics of the horizon formation,
deferred for later studies.  
When this paper were near-completed, we learned about an 
interesting work by Hubeny,Liu and Rangamani \cite{hubeny}
who discuss certain
null geodesics related observation points at the boundary.
Since phases of the waves add coherently near geometric optics paths
(null geodesics), this is another interesting form of coherence,
although perhaps unrelated to our WKB  ``echos''.

Finally, one may now think about relaxing our main assumption --
flatness of the shell, e.g. by including first corrections
resulting from slow variations of its position. In this case
the metric above the shell would become  time-dependent,
allowing a ``single point observer'' to see some 
relaxation dynamics as well.

\section{Acknowledgment}
The work is partially
supported by the US-DOE grants DE-FG02-88ER40388 and
DE-FG03-97ER4014. S.L. would like to thank Keun-young Kim, Peng Dai,
Yu-tin Huang and Elli Pomoni for multiple discussions.
\appendix

\section{WKB treatment of (\ref{Z_eom})}\label{wkb}

In order to apply WKB method to (\ref{Z_eom}), we need to convert it Schrodinger-
type equation. Introducing a new field $\psi=\sqrt{\frac{1-u^2}{u}}Z_3$,
(\ref{Z_eom}) becomes:

\be\label{schrodinger}
\psi''+\frac{\omg^2(1-s^2(1-u^2))}{u(1-u^2)^2}\psi+\frac{-3+6u^2+u^4}{4u^2(1-u^2)^2}\psi=0
\ee

with $s=\lv\frac{q}{\omg}\rv\approx 1$. $\omg$ is a large parameter to
justify WKB. There are two singularities in (\ref{schrodinger}): $u=0$,$u=1$.
The term proportional to $\omg^2$ may vanish at $u_0=\frac{\sqrt{s^2-1}}{s}$
if $s>1$. We discuss two cases separately: I $s<1$, II $s>1$ and focus on
$\omg>0$ only. The solution for $\omg<0$ can be obtained easily from the
solution for $\omg>0$ by the substitution $c_+ \leftrightarrow c_-$.

Case I $s<1$: Away from the singularities, the WKB 
solution to (\ref{schrodinger}) is given as:

\be
\psi_\pm=S'^{-1/2}e^{\pm i\omg\int_0^u S'du}
\ee

with $S'=\sqrt{\frac{1-s^2(1-u^2)}{u(1-u^2)^2}}$

Near the singularities, (\ref{schrodinger}) becomes:

\be
&&u\rightarrow0 \no
&&\psi''+\frac{\omg^2(1-s^2)}{u}\psi-\frac{3}{4u^2}\psi=0 \Rightarrow 
\psi=\sqrt{u}H_2^{(1),(2)}(2\omega\sqrt{1-s^2}\sqrt{u}) \no
&&u\rightarrow1 \no
&&\psi''+\frac{1+\omg^2}{4(1-u)^2}\psi=0 \Rightarrow
\psi=(1-u)^{\frac{1\pm i\omg}{2}}
\ee

On the other hand, we have:

\be
&&u\rightarrow0 \quad S'\rightarrow\sqrt{\frac{1-s^2}{u}} \int_0^u S'du=2\sqrt{u}\sqrt{1-s^2} \no
&&u\rightarrow1 \quad S'\rightarrow\frac{1}{2(1-u)} \int_0^u S'du=a_0-\frac{1}{2}ln(1-u)
\ee

Matching the WKB solution with the approximate solutions near the singularities
(using asymptotic expansion of Hankel function), we obtain:

\be
&&\psi=c_+(1-u)^{\frac{1+i\omg}{2}}+c_-(1-u)^{\frac{1-i\omg}{2}} \no
&&\sim c_+e^{i\omg a_0}\psi_-+c_-e^{-i\omg a_0}\psi_+ \no
&&\sim c_+e^{i(\omg a_0-\frac{5\pi}{4})}\sqrt{u}H_2^{(2)}(2\omega\sqrt{1-s^2}\sqrt{u})+c_-e^{-i(\omg a_0-\frac{5\pi}{4})}\sqrt{u}H_2^{(1)}(2\omega\sqrt{1-s^2}\sqrt{u}) \no
&&\sim\sqrt{u}\(H_2^{(2)}(2\omg\sqrt{1-s^2}\sqrt{u})+\frac{ic_-}{c_+}e^{-2i\omg a_0}H_2^{(1)}(2\omg\sqrt{1-s^2}\sqrt{u}\)
\ee

Case.II $s>1$: This case is a little more complicated because WKB approximation
breaks down near $u=u_0$. Away from $u=0,1,u_0$, we have the following WKB
solutions:

\be
u>u_0 \quad \psi_\pm^>=S'^{-1/2}e^{\pm i\omg\int_{u_0}^u S'du} \no
u<u_0 \quad \psi_\pm^<=S'^{-1/2}e^{\pm\omg\int_u^{u_0}{\bar S'}du}
\ee

where $S'=\sqrt{\frac{1-s^2(1-u^2)}{u(1-u^2)^2}}$,${\bar S'}=\sqrt{\frac{s^2(1-u^2)-1}{u(1-u^2)^2}}$.

We first match WKB solutions at $u=u_0$. Near $u=u_0$, (\ref{schrodinger})
becomes:

\be\label{airy}
\psi''+\omg^2a(u-u_0)\psi+b\psi=0
\ee

where $a=\frac{2s^2u_0}{u_0(1-u_0^2)^2}$,$b=\frac{-3+6u_0^2+u_0^4}{4u_0^2(1-u_0^2)^2}$

(\ref{airy}) can be solved by Airy functions:

\be
\psi=Ai(-\frac{\omg^2a(u-u_0)+b}{(\omg^2a)^{2/3}}) \no
\psi=Bi(-\frac{\omg^2a(u-u_0)+b}{(\omg^2a)^{2/3}})
\ee

Using the asymptotic expansion of Airy functions:$(x>0)$

\be
&&Ai(x)\sim\frac{e^{-\frac{2}{3}x^{3/2}}}{2\sqrt{\pi}x^{1/4}} \no
&&Bi(x)\sim\frac{e^{\frac{2}{3}x^{3/2}}}{\sqrt{\pi}x^{1/4}} \no
&&Ai(-x)\sim\frac{sin(\frac{2}{3}x^{3/2}+\frac{1}{4}\pi)}{\sqrt{\pi}x^{1/4}} \no
&&Bi(x)\sim\frac{cos(\frac{2}{3}x^{3/2}+\frac{1}{4}\pi)}{\sqrt{\pi}x^{1/4}}
\ee

We obtain the following match between WKB solutions:

\be
\frac{C}{2}(\psi_+^>+\psi_-^>)+\frac{D}{2i}(\psi_+^>-\psi_-^>)\sim
\frac{C-D}{2}\psi_+^<+\frac{C+D}{4}\psi_-^<
\ee

Next we match WKB solutions with approximate solutions near singularities
similarly as case I. Finally we have:

\be
&&\psi=c_+(1-u)^{\frac{1+i\omg}{2}}+c_-(1-u)^{\frac{1-i\omg}{2}} \no
&&\sim\(c_+e^{i\omg b_0}-ic_-e^{-i\omg b_0}\)\frac{e^{\omg c_0}}{\pi}\sqrt{u}K_2(2\omg\sqrt{s^2-1}\sqrt{u}) \no
&&+\frac{1}{2}\(-ic_+e^{i\omg b_0}+c_-e^{-i\omg b_0}\)e^{-\omg c_0}\sqrt{u}I_2(2\omg\sqrt{s^2-1}\sqrt{u}) \no
&&\sim\sqrt{u}\(1-\frac{ic_-}{c_+}e^{-2i\omg b_0}\)\frac{e^{\omg c_0}}{\pi}K_2(2\omg\sqrt{s^2-1}\sqrt{u}) \no
&&-\frac{i\sqrt{u}}{2}\(1+\frac{ic_-}{c_+}e^{-2i\omg b_0}\)e^{-\omg c_0}I_2(2\omg\sqrt{s^2-1}\sqrt{u})
\ee

with $lim_{u\rightarrow 1}\int_{u_0}^u S'du=b_0-\frac{1}{2}ln(1-u)$,$\int_0^{u_0}{\bar S'}du=c_0$

Summarizing all cases, we have:

\be
\psi=\left\{\begin{array}{l@{\quad}l}
\sqrt{u}\(H_2^{(2)}(2\sqrt{\omg^2-q^2}\sqrt{u})
+\frac{ic_-}{c_+}e^{-2i\omg a_0}H_2^{(1)}(2\sqrt{\omg^2-q^2}\sqrt{u})\)& \omg>\lv q\rv\\
\sqrt{u}\(H_2^{(1)}(2\sqrt{\omg^2-q^2}\sqrt{u})
-\frac{ic_-}{c_+}e^{-2i\omg a_0}H_2^{(2)}(2\sqrt{\omg^2-q^2}\sqrt{u})\)& \omg<-\lv q\rv\\
\sqrt{u}\(1-\frac{ic_-}{c_+}e^{-2i\omg b_0}\)\frac{e^{\omg c_0}}{\pi}K_2(2\sqrt{q^2-\omg^2}\sqrt{u})& \\
-\frac{i\sqrt{u}}{2}\(1+\frac{ic_-}{c_+}e^{-2i\omg b_0}\)e^{-\omg c_0}I_2(2\sqrt{q^2-\omg^2}\sqrt{u})& 0<\omg<\lv q\rv\\
\sqrt{u}\(1+\frac{ic_-}{c_+}e^{-2i\omg b_0}\)\frac{e^{\omg c_0}}{\pi}K_2(2\sqrt{q^2-\omg^2}\sqrt{u})& \\
+\frac{i\sqrt{u}}{2}\(1-\frac{ic_-}{c_+}e^{-2i\omg b_0}\)e^{-\omg c_0}I_2(2\sqrt{q^2-\omg^2}\sqrt{u})& 0>\omg>-\lv q\rv
\end{array}
\right.
\ee

\section{Gauge Choice for the Sound Channel}\label{gc}

The aim of the section is to show it is possible render the following
matching condition by a proper choice of gauge:

\be\label{mc_gc}
&&\phi_2^f=\phi_2 \no
&&\phi_2^f{'}=\frac{1}{\sqrt{f}}\phi_2{'}+\frac{\ka_5^2 p}{3u}\frac{\phi_2{'}}{\sqrt{f}}
\ee

With the help of (\ref{mc_p}), (\ref{mc_gc}) can be simplified to:

\be
&&\wh_{aa}-\frac{2-f}{\sqrt{f}}h_{aa}^f=0 \no
&&\wh_{aa}{'}+\frac{\ka_5^2 p}{3u}\wh_{aa}-(2-f)\wh_{aa}^f{'}+f'h_{aa}^f=0
\ee

where all the quantities are evaluated at $u=u_m$.

We note the residue gauge degree of freedom implies that it is sufficient
to satisfy (\ref{mc_p}) up to a gauge choice. In particular, we could add
pure gauge solution to the sound channel:
$h^{gauge}$ inside the shell and $h^{gauge,f}$ outside.
According to \cite{sound}, the only pure gauge solution that touch 
$h_{aa}^f$($h_{aa}$ with $f=1$) is:

\be
&&h_{tt}^{gauge,f}=\frac{\sqrt{f}(1+u^2+2\omg^2u)}{u} \no
&&h_{tw}^{gauge,f}=\frac{-q\omg\arcsin u-q\omg u\sqrt{f}}{u} \no
&&h_{aa}^{gauge,f}=-\frac{2\sqrt{f}}{u} \no
&&h_{ww}^{gauge,f}=\frac{2q^2\arcsin u-\sqrt{f}}{u}
\ee

Now, it is enough to satisfy:

\be\label{mc_gc2}
&&{\bar h}_{aa}-\frac{2-f}{\sqrt{f}}{\bar h}_{aa}^f=0 \no
&&{\bar h}_{aa}'+\frac{\ka_5^2 p}{3u}{\bar h}_{aa}-(2-f){\bar h}_{aa}^f{'}
+f{'}{\bar h}_{aa}^f=0
\ee

where ${\bar h}_{aa}=\wh_{aa}+\frac{A}{u}$ and 
${\bar h}_{aa}^f=\wh_{aa}^f+\frac{B\sqrt{f}}{u}$.
It is easy to see it is always possible to satisfy (\ref{mc_gc2}) with a
proper choice of constants $A$ and $B$.

\end{document}